\documentclass[a4paper,final,showpacs]{revtex4-1}

\usepackage{graphicx}
\usepackage{gensymb}
\usepackage{Citesort}
\usepackage{color}

\usepackage{epsfig}
\usepackage{amssymb}
\usepackage{amsmath}
\usepackage{booktabs}
\usepackage{siunitx}
\usepackage{datetime}

\usepackage{geometry}
\usepackage{mathdots}
\usepackage{MnSymbol}

\begin{document}

\title[Radiative and collisional processes in translationally cold samples of hydrogen Rydberg atoms]{Radiative and collisional processes in translationally cold samples of hydrogen Rydberg atoms studied in an electrostatic trap}
\author{Ch.~Seiler$^1$, J.~A.~Agner$^1$, P.~Pillet$^2$, and F.~Merkt$^1$}
\affiliation{$^1$ Laboratorium f\"ur Physikalische Chemie, ETH Z\"urich, 8093~Z\"urich, Switzerland \\
$^2$ Laboratoire Aim\'{e} Cotton, CNRS, Universit\'{e} Paris-Sud, ENS~Cachan, 91405~Orsay, France}

\begin{abstract}
Supersonic beams of hydrogen atoms, prepared selectively in Rydberg-Stark states of principal quantum number $n$ in the range between 25 and 35, have been deflected by 90$^\circ$, decelerated and loaded into off-axis electric traps at initial densities of $\approx 10^6$ atoms/cm$^{-3}$ and translational temperatures of 150 mK. The ability to confine the atoms spatially was exploited to study their decay by radiative and collisional processes. The evolution of the population of trapped atoms was measured for several milliseconds in dependence of the principal quantum number of the initially prepared states, the initial Rydberg-atom density in the trap, and the temperature of the environment of the trap, which could be varied between 7.5 K and 300 K using a cryorefrigerator. At room temperature, the population of trapped Rydberg atoms was found to decay faster than expected on the basis of their natural lifetimes, primarily because of absorption and emission   stimulated by the thermal radiation field. At the lowest temperatures investigated experimentally, the decay was found to be multiexponential, with an initial rate scaling as $n^{-4}$ and corresponding closely to the natural lifetimes of the initially prepared Rydberg-Stark states. The decay rate was found to continually decrease over time and to reach an almost $n$-independent rate of more than (1\,ms)$^{-1}$ after  3\,ms. To analyze the experimentally observed decay of the populations of trapped atoms, numerical simulations were performed which included all radiative processes, i.e., spontaneous emission as well as absorption and emission stimulated by the thermal radiation. These simulations, however, systematically underestimated the population of trapped atoms observed after several milliseconds by almost two orders of magnitude, although they reliably predicted the decay rates of the remaining atoms in the trap. The calculations revealed that the atoms that remain in the trap for the longest times have larger absolute values of the magnetic quantum number $m$ than the optically prepared Rydberg-Stark states, and this observation led to the conclusion that a much more efficient mechanism than a purely radiative one must exist to induce transitions to  Rydberg-Stark states of higher $\left|m\right|$ values. While searching for such a mechanism, we discovered that resonant dipole-dipole collisions between Rydberg atoms in the trap represent an extremely efficient way of inducing transitions to states of higher $\left|m\right|$ values. The efficiency of the mechanism is a consequence of the almost perfectly linear nature of the Stark effect at the moderate field strengths used to trap the atoms, which permits cascades of transitions between entire networks of near-degenerate Rydberg-atom-pair states. To include such cascades of resonant dipole-dipole transitions in the numerical simulations, we have generalized the two-state
F\"orster-type collision model used to describe resonant collisions in ultracold Rydberg gases to a multi-state situation. It is only when considering the combined effects of collisional and radiative processes that
the observed decay of the population of Rydberg atoms in the trap could be satisfactorily reproduced for all $n$ values studied experimentally.
\end{abstract}
\maketitle


\section{\label{sc:introduction}Introduction}

Rydberg states of atoms and molecules are electronically excited states the spectral positions $E_n$ of which can be described in good approximation by Rydberg's formula:
\begin{equation} \label{Rydberg}
E_n=E_{\rm I} - hcR_{\rm M}/(n-\delta_\ell)^2,
\end{equation}
in which $R_{\rm M}$ represents the mass-dependent Rydberg constant, $\delta_\ell$ the quantum defect, and $n$ the principal quantum number.
For each value of the orbital-angular-momentum quantum number $\ell$, Rydberg states form infinite series, which converge for $n\rightarrow \infty$ to one of the ionization limits ($E_{\rm I}$ in Eq.~(\ref{Rydberg})) of the atom or molecule. The physical properties of Rydberg states typically scale as integer powers of $n$ and thus strongly depend on its values \cite{bethe57a,gallagher94a,merkt97a}. The properties of Rydberg states also depend on the orbital and magnetic quantum numbers $\ell$ and $m$. The selective production of Rydberg states of well-defined quantum numbers enables one to generate atoms or molecules with specific physical properties. This advantage is exploited in an increasing number of scientific applications in several subfields of physics and chemistry, such as quantum optics (see, e.g., \cite{raimond01a,haroche13a}), quantum-information science (see, e.g., \cite{gaetan09a,saffman10a}), metrology in atoms and molecules (see, e.g.,  \cite{biraben09a,wall15a,sprecher11a}), high-resolution photoelectron spectroscopy \cite{muellerdethlefs98a}, the sensing of electromagnetic fields \cite{frey93a,osterwalder99a,thiele14a}, and in research on antihydrogen \cite{gabrielse12a,amoretti02a,kellerbauer08a}.

In the laboratory, Rydberg atoms and molecules are usually formed by photoexcitation from lower-lying electronic states and refined schemes enable one to selectively access states of almost any values of the quantum numbers $n$, $\ell$ and $m$, remarkable examples being the production of states of very high principal quantum number \cite{ye13a} or states of maximal $m$ value \cite{delande88a,hare88a}. Rydberg states are also formed by recombination of an electron and a cation in any environment where electrons and cations are present, such as in plasmas or in ionized regions of the interstellar medium.

Photoexcitation from the ground state or low-lying electronic states typically provides access to penetrating Rydberg states, i.e. states with low-$\ell$ and $m$ character, which, in atoms, decay radiatively with lifetimes scaling as $n^3$ or $n^4$ depending on whether the excitation is carried out under field-free conditions or not and whether the Stark effect is strong enough to induce $\ell$ mixing or not. In molecules, Rydberg states of low-$\ell$ character usually decay by internal conversion, predissociation or autoionization because these processes typically have much higher rates than radiative processes. In contrast, recombination preferentially produces Rydberg states of high-$\ell$ and $m$ character, which have much longer lifetimes and primarily decay through sequences of low-frequency radiative transitions to Rydberg states of neighboring $n$, $\ell$ and $m$ values.

The radiative properties of high Rydberg states are very different from those of low-lying electronic states because of the electric-dipole transition moments that connect states of similar quantum numbers scale as $n^2$ and the corresponding transition frequencies scales as $n^{-3}$. Blackbody radiation can therefore efficiently stimulate emission or absorption processes and, in some cases, even can be the primary source of decay. The influence of blackbody radiation on the behavior of Rydberg states is well known and has been extensively investigated experimentally and theoretically  \cite{gallagher79a,spencer82a,fabre82a,lehman83a,filipovicz85a,gallagher94a,qiu99a,flannery03a,pohl06a,robicheaux06a,glukhov10a,mack15a}.

Some of the scientific applications listed above exploit specific physical properties of Rydberg states, e.g. their large polarizability $\alpha$ for the sensing of electric fields ($\alpha$ scales as $n^7$ \cite{gallagher94a}), the large van der Waals interaction between Rydberg atoms (the $C_6$ coefficient scales as $n^{11}$) to generate quantum gates, or their long lifetimes, which is essential for applications in high-resolution photoelectron spectroscopy. Other applications are concerned with the formation of Rydberg states by recombination and, in the case of research on antihydrogen, would benefit from a rapid conversion of the Rydberg states formed in the recombination process to the ground state  \cite{gabrielse12a}. Transitions stimulated by thermal radiation can modify the desired properties and must  be considered in the planning of the experiments, and sometimes also in the analysis of the results. Successive transitions induced by thermal radiation leads to the gradual redistribution of a population of Rydberg states to a wide range of states of different $n$, $\ell$ and $m$ values. Some of these states then rapidly fluoresce to the ground state or low-lying electronic states. Others can be ionized by the thermal radiation. Overall, it is difficult to reliably predict the evolution of a population of Rydberg atoms over long time scales. Predictions become even more difficult if collisions with atoms and molecules in the background gas or with other Rydberg atoms also contribute to induce transitions. In discussions related to the planning of experiments on antihydrogen in the realm of the AEGIS collaboration \cite{kellerbauer08a}, we became aware that it is not straightforward to reliably predict the fate of an initially prepared sample of Rydberg atoms, in particular the time needed for the ground state to be populated and the yield of ground-state atoms at a given time.

In the past ten years, we have contributed to the development of methods with which translational cold samples of Rydberg atoms and molecules in supersonic beams can be decelerated and stored in electric traps \cite{vliegen04a,vliegen07a,hogan08a,hogan09a}. To obtain experimental information on the evolution of an initial population of Rydberg atoms and molecules selectively prepared in Rydberg-Stark states, we have also measured the decay of the population of trapped Rydberg atoms and molecules for several hundreds of microseconds \cite{hogan08a,seiler11a,seiler11b,seiler12a}.
These measurements are complementary to recent all-optical measurements of the lifetimes of selected $nl$ Rydberg states following excitation of ultracold atoms in MOT, which reveal the decay of the initially prepared Rydberg states of Rb with principal quantum numbers in the range 30-40 by fluorescence, blackbody-radiation-induced and collisional processes \cite{mack15a} and to earlier studies of the influence of blackbody radiation, in particular ionization, on the lifetimes and other properties of Rydberg states of atoms (see, e.g., Refs. \cite{gallagher79a,spencer82a,fabre82a,lehman83a,filipovicz85a,qiu99a,flannery03a,pohl06a,robicheaux06a}, and \cite{gallagher94a} and references therein)
In an effort to understand how the different radiative or collisional processes affect the evolution of Rydberg atom population and induce losses of atoms and molecules from the traps, we have successively extended the initial deceleration and trapping experiments by (i) deflecting the Rydberg atoms and molecules from the beam axis before loading them into off-axis traps in order to suppress collisions with other atoms and molecules in the beam \cite{seiler11a}, (ii) installing successive thermal shields around the traps and thermalizing these shields to progressively lower temperatures, down to below 10 K, (iii) comparing the behavior of atomic and molecular Rydberg states in experiments carried out on H atoms and H$_2$ molecules \cite{seiler11b}, and (iv) trying to model the relevant radiative and collisional processes.

We now realize that such an effort never reaches the stage where the experiments and their analysis can be considered to be entirely satisfactory. At the same time, we also feel that the results obtained so far, despite their incompleteness and imperfections, should be made available. We present here the results we have obtained with atomic hydrogen, and will report our results on molecular hydrogen in a future publication. Atomic hydrogen was of particular interest in our procedure, for several reasons. Firstly, its Rydberg states all have zero quantum defects, their Stark effect is to a good approximation linear at low fields, and their electric-field-ionization dynamics is diabatic \cite{gallagher94a,vliegen06b}. Secondly, the decay dynamics of isolated H Rydberg atoms are entirely governed by radiative processes and are not affected by molecular processes such as predissociation and autoionization. Thirdly, analytic expressions can be used to calculate the rates of spontaneous and stimulated radiative processes in H-atom Rydberg states. Finally, the results obtained on hydrogen can immediately be transferred to antihydrogen, which was the main motivation of our original experiments on H atoms.

This article is structured as follows: The experimental apparatus and procedure are described in Section~\ref{sc:experiment}. The experimental results are then presented in Section~\ref{sc:results}. Section~\ref{sc:lifetime} summarizes the expressions we have used to calculate the rates of spontaneous and stimulated radiative processes and describes the Monte-Carlo-simulation procedure we followed to calculate the loss of atoms from the trap caused by radiative processes. The role of collisions is then discussed in Section~\ref{sec:ch6-theory}. This section also presents a simple resonant-dipole-dipole-collision model, which explains the aspects of the experimental data that could not be accounted for by radiative processes only. The article ends with a short conclusions section.


\section{\label{sc:experiment}Experiment}

\subsection{\label{subsc:laser-setup} Laser system, excitation scheme, and Rydberg-Stark decelerator and trap}

\begin{figure*}[!tbh]
\centering
\includegraphics[angle=0,width=1.0\textwidth]{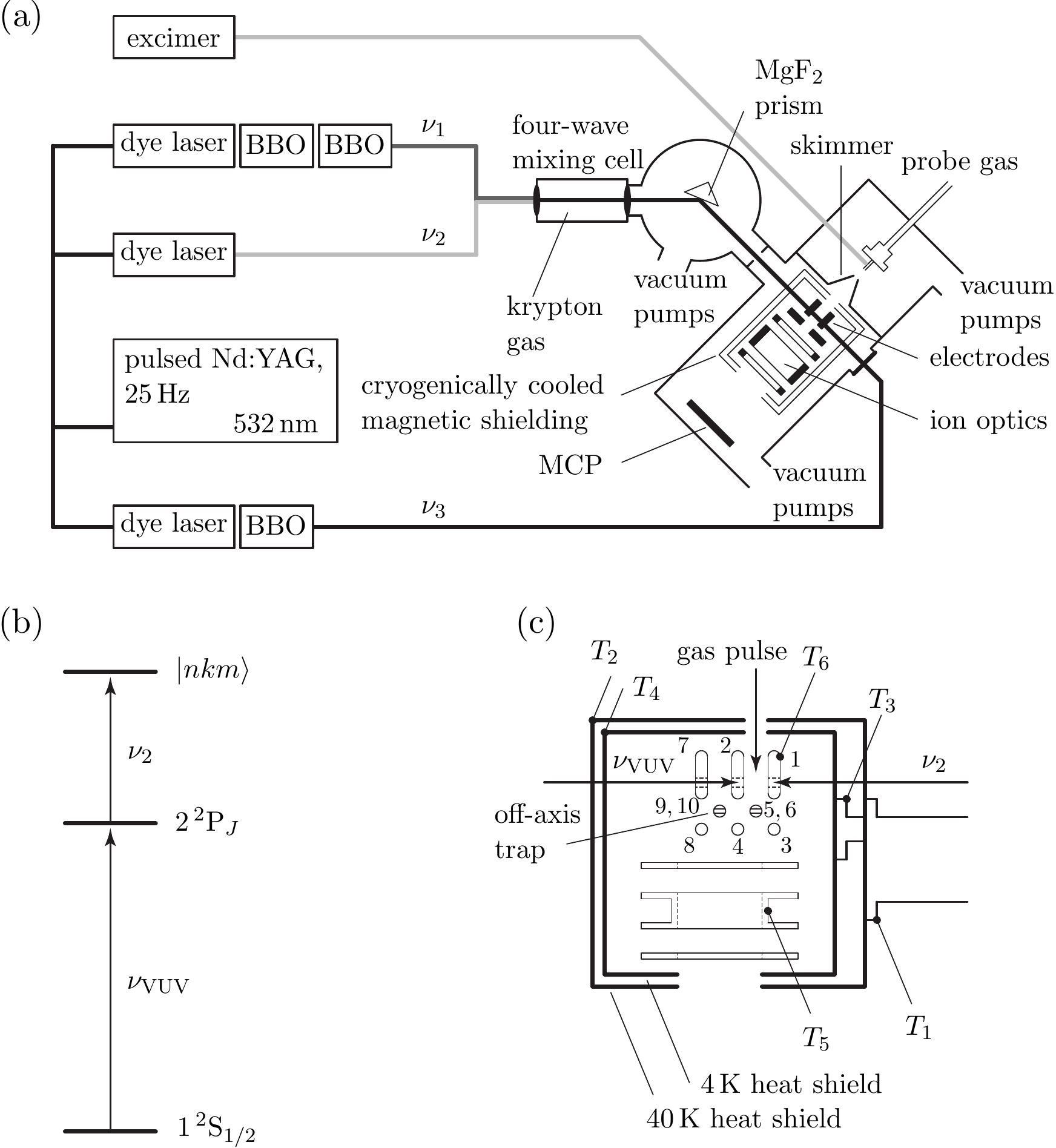}
\caption{\label{fig:h-setup} (a) Schematic representation of the laser setup and vacuum chambers employed in the experiments on atomic hydrogen. (b)~Excitation sequence used to populate Rydberg states of atomic hydrogen. (c) Off-axis Rydberg-Stark decelerator depicting the deceleration region, the three plates of the ion lens, and the two heat shields connected to the pulse-tube cooler. T$_1$ to T$_6$ indicate the locations of the temperature sensors.}
\end{figure*}

Figure~\ref{fig:h-setup}(b) depicts the two-photon excitation scheme employed to populate the Rydberg states of atomic hydrogen, as described in detail in Refs.~\cite{vliegen06b,seiler11a}. Pulsed coherent vacuum-ultraviolet radiation at Lyman-$\alpha$ (121.6\,nm), generated by four-wave mixing in Kr, was used to excite the hydrogen atoms from the $1\,^2\text{S}_{1/2}$ ground state to the $2\,^2\text{P}_J$ intermediate state. A second pulsed laser with tunable frequency was used to excite the hydrogen atoms from the intermediate state to $\left|nkm=0,\pm 2\right>$ Rydberg-Stark states suitable to deceleration and trapping.

A schematic representation of the apparatus used to prepare Rydberg-Stark states of atomic hydrogen is displayed in Figure~\ref{fig:h-setup}(a). The experimental setup consists of (i)~the laser system, with three dye lasers pumped at a repetition rate of 25 Hz by the doubled (532 nm) and tripled (355 nm) outputs of a Nd:YAG laser, a four-wave mixing cell filled with Kr at a pressure in the range 20-40 mbar, and optical elements required to frequency double and recombine the different laser beams (see Ref.~\cite{vliegen06b,seiler11a} for details), and (ii)~a set of differentially pumped vacuum chambers including a monochromator chamber, a gas-source chamber and a photoexcitation, deceleration and trapping chamber. This last chamber also contains a time-of-flight spectrometer with a microchannel-plate (MCP) detector used to monitor the H$^+$ ions produced by pulsed field ionization of the trapped Rydberg atoms. Because of the 0.7~cm$^{-1}$ bandwidth of the radiation used to excite the Rydberg-Stark state from the intermediate level and the inhomogeneity of the electric field distribution in the excitation region, it was not possible to prepare individual Stark states, but typically several Rydberg-Stark states located energetically close to the desired Stark states and having similar electric dipole moments. The perpendicular arrangement of the polarization of both lasers with respect to the dc electric field in the photoexcitation region led to a dominant excitation of $\left|m\right|=2$ Stark states (82 \%) with a weaker contribution of $m=0$ states (18 \%). The Stark effect in Rydberg states of H (see Ref.~\cite{gallagher94a} and also Fig.~\ref{fig:stark-levels}a below) is such that the optically accessible $m=0,\pm 2$ Stark states have even (odd) $k$ values for odd (even) values of $n$. The range of $k$ states was adjusted from 18-26 at $n=27$ to 14-22 at $n=33$, so that the average value of the electric dipole moment was approximately the same in all experiments, i.e., about 900 $a_0e$,  which is ideal for deceleration and trapping \cite{vliegen06b,hogan08a,seiler11a}.

Hydrogen atoms were produced by pulsed excimer-laser photolysis of NH$_3$ at 193\,nm (ArF) in a quartz capillary (1\,mm inside diameter, $\approx 15$\,mm length) mounted on the base plate of the pulsed valve, using a method described in Ref.~\cite{willitsch04b} and later used to generate supersonic beams of atomic hydrogen~\cite{vliegen06b}. The supersonic beam formed at the exit of the capillary had a mean longitudinal velocity of $\sim 500$\,m/s ($\sim 600$\,m/s) when Kr (Ar) was used as carrier gas for the supersonic expansion.

The Rydberg-Stark decelerator and trap, consisting of 10 electrodes and depicted schematically in Fig.~\ref{fig:h-setup}(c), was built of copper and the surfaces were coated with a thin ($\sim 0.25\,\mu$m) gold layer. Because this setup was designed to be cooled down to temperatures below 10~K, all electrodes were thermally connected with each other and anchored to the cold plate of a pulse-tube cooler using 1-mm-thick alumina (Al$_2$O$_3$) pads. Small nickel-coated plugs were directly soldered to each electrode, enabling the straightforward connection of the cables used to apply time-dependent electric potentials on the electrodes for deceleration and trapping of the H Rydberg atoms.

Sequences of pulsed electric potentials were applied for deceleration, 90$^\circ$ deflection and storage of the Rydberg atoms in a three-dimensional electric trap displaced laterally by 6 mm from the original supersonic-beam propagation axis, as described in Ref.~\cite{seiler11a}. The
pulse sequence described in Ref.~\cite{seiler11a} (see their Fig.~2) was only very slightly modified to take into account small differences in the initial longitudinal velocity of the supersonic beam.
During the photoexcitation and trapping phases of the experiment, electric potentials of $+12$\,V and $-12$\,V were applied to electrodes 1, 4 and 7, and 2, 3 and 8 , respectively (see Fig.~\ref{fig:h-setup}(c) and Fig.~1 in Ref.~\cite{seiler11a}), resulting in an electric-field strength of 58\,V/cm at the position of photoexcitation and in a two-dimensional off-axis quadrupolar electric trap used to store the Rydberg atoms. Confinement in the third dimension was achieved by applying potentials of $-22$\,V ($+22$\,V) to end-cap electrodes 5 and 9 (6 and 10). These potentials led to an electric field of 10\,V/cm at the trap minimum and a saddle point in the electric-field distribution located 36\,V/cm above the field strength at the trap minimum, which corresponds to a trap depth of 1.3\,cm$^{-1}$ (or $E/k_{\text{B}}=1.9\,$K) for an $n=30$, $k=19$ hydrogenic Rydberg-Stark state. The translational temperature of the H Rydberg-atom sample after deceleration and trapping was about 150 mK ~\cite{seiler11a}. All electrodes used for deceleration, field ionization and ion extraction were placed inside a heat shield to reduce scattered thermal radiation.

Rydberg hydrogen atoms loaded into the off-axis trap were detected after an adjustable delay by pulsed field ionization (PFI) with a pulsed potential of $\approx 1.5$\,kV applied to electrodes 2 and 7. This pulsed potential also extracted the protons toward the MCP detector. An einzel lens was used to reduce the spherical and chromatic aberrations of the ion optics compared to the previous version employed in the experiments presented in Ref.~\cite{seiler11a}.

In all experiments described below, the trap-loss processes were measured by monitoring the H$^+$ ions produced by PFI of the trapped Rydberg atoms as a function of the time delay between photoexcitation and the electric-field pulse. In the following, we refer to the results of such measurements as trap-decay curves. These curves were obtained in subsequent experimental cycles consisting each of a photoexcitation, deceleration, trapping, field ionization, and  detection phase.
To obtain a trap-decay curve, the delay between photoexcitation and PFI was gradually increased until the field ionization signal dropped below the detection threshold.  Such a curve only contains indirect information on the lifetimes of the Rydberg-Stark states prepared optically. Indeed, transitions induced by blackbody radiation or collisions do not always lead to a loss of atoms from the trap or to a reduction of the PFI signal because the final states of the transitions can be Rydberg-Stark states that have (i) a dipole moment sufficiently high for the atoms to remain trapped and (ii) a principal quantum number sufficiently high to be field ionized, i.e., $n \ge 23$.

Disentangling the different radiative and collisional processes responsible for the observed trap-decay curves represented a challenge and necessitated systematic studies of Rydberg-Stark states of different principal quantum numbers, carried out at different blackbody-radiation temperatures and different initial Rydberg-atom densities. Trapping the atoms off axis suppressed the undesirable effects of collisions between the trapped Rydberg atoms with atoms and  molecules in the trailing part of the gas pulses, as explained and demonstrated in Ref.~\cite{seiler11a}. Cooling the entire decelerator and trapping region to temperatures down to 10 K was essential in quantifying the effects of blackbody radiation. Reducing the initial Rydberg-atom density turned out to be difficult, because the resulting decrease of signal strongly shortened the time window over which the trap decay curves could be measured. Fortunately, trap-loss processes naturally reduced the density of trapped Rydberg atoms with time so that density-dependent effects could be inferred by numerical simulation of the trap-decay curves.


\subsection{\label{subsc:cryo-setup} Cryogenic decelerator setup}

To minimize stray magnetic fields and to investigate the effects of blackbody radiation, the Rydberg-Stark decelerator and trapping device was enclosed in a set of concentric magnetic and thermal shields consisting of an inner, combined magnetic and thermal shield (0.5-mm-thick Cryoperm, coated with a 100-$\mu$m-thick Ag layer), an outer thermal shield made of 1-mm-thick copper, and an outer mu-metal shield made of 1-mm-thick Permalloy. The outer and inner thermal shields, referred to below as the 40\,K and 4\,K heat shields, were connected to the 40\,K and 4\,K temperature stages, respectively, of a pulse-tube cryorefrigerator (\textsc{Cryomech}, PT415).  The specified cooling power of the 4\,K and 40\,K stages were 1.35\,W at 4.2\,K and 36\,W at $\approx 45$\,K, respectively.
The 40\,K heat shield, which was aligned with the main experimental chamber by four tapered dove tails fabricated from a polyallylamid-glass compound (\textsc{Solvay Plastics}, IXEF 1022), served the purpose of shielding the inner region from room-temperature radiation and made it possible to reach low temperatures inside the inner heat shield.

 To minimize the thermal load on the cold stage of the pulse-tube cooler, all copper cables and wires inside the vacuum chamber, which were initially used because of their excellent electric and thermal conductivity, were replaced by phosphor-bronze wires (made from a CuSnP alloy, \textsc{LakeShore}, single strand cryogenic wire, SL-32). These wires are nonmagnetic and do not efficiently transfer heat. All cables were thermalized by connection, via micro-plugs, to the surface electrodes of thin circuit boards printed on polymer foils glued to small copper blocks. The copper blocks were mounted onto the 40\,K and 4\,K base plates, which were themselves attached to the ends of the pulse tube cooler via $100\,\mu$m thick indium foils using thermally conducting grease (\textsc{Apiezon}~N).

Six silicon-based temperature sensors (\textsc{LakeShore}, DT-670), placed at the locations marked T$_1$-T$_6$ in Fig.~\ref{fig:h-setup}(c), were used to monitor the temperature at the 40\,K base plate of the pulse-tube cooler (T$_1$), at the 40\,K (T$_2$) and 4\,K (T$_3$) mounting plates, at the 4\,K heat shield (T$_4$), at the bottom of the middle electrode of the einzel lens (T$_5$), and at one of the deceleration electrodes (T$_6$). The four cables (\textsc{LakeShore}, quad-lead cryogenic wire, QL-32) of each of the six temperature sensors were also thermally anchored using the technique described above. To adjust the temperature of the deceleration and trapping region between 7.5\,K and 25\,K, heating resistors were mounted at the 4\,K and 40\,K stages.

A small amount of room-temperature radiation inevitably enters the inner region through the holes used for the supersonic beam, the laser beams and the ion time-of-flight spectrometer, resulting in a temperature of 11\,K in the inner region, as measured with temperature sensor T$_6$. The room-temperature contribution to the radiation field penetrating inside the 40\,K heat shield  was determined from the construction drawings to correspond to a solid angle of $\approx 1\%$. To further reduce this solid angle and reach temperatures as low as 7.5\,K in the inner region, baffles with a slit aperture were inserted at the major holes between the 40\,K heat shield and the room-temperature environment.

Additional measures were taken to verify that room-temperature radiation penetrating through the holes of the heat shields did not significantly affect the measurements. To avoid direct sight from the MCP detector to the trap region, a 90$^\circ$ deflection stage for the H$^+$ ion beam was inserted, which enabled us to block the largest hole with a plate connected to the 40\,K heat shield. To minimize reflection of thermal radiation within the inner shield, a third shield coated with microwave-absorbing epoxy (Eccosorb) was installed around the trap inside the 4\,K heat shield.
Although the presence of metallic reflectors and a residual direct exposure to room-temperature environment still leaves the possibility of wavelength-dependent, spatially inhomogeneous warm contribution to the radiation field in the Rydberg-atom trap, the measures described above did not significantly affect the observed trap-decay curves. We therefore conclude that the room-temperature contribution to the radiation field in the Rydberg-atom trap is effectively reduced to $1\%$.


\section{\label{sc:results}Experimental Results}
\begin{figure*}[!tbh]
\centering
\includegraphics[angle=270,width=1.0\textwidth]{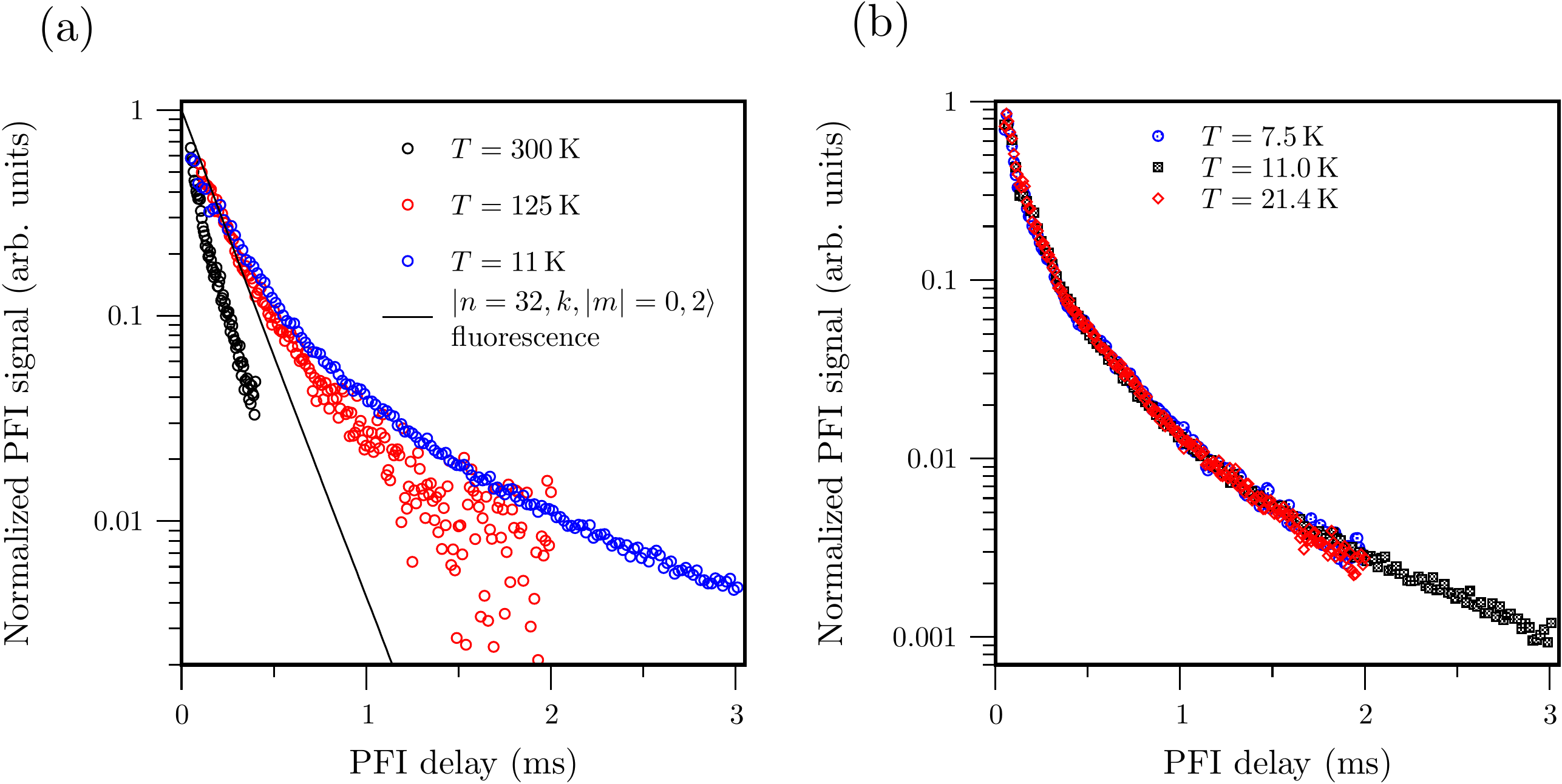}
\caption{\label{fig:ch6-temperature}(a)~Trap-decay curves measured following initial excitation and trapping of  $\left|n=32,k,\left|m\right|=0,\pm 2\right>$ Rydberg-Stark states with $k$ in the range $15-23$ at room temperature (black circles) and at blackbody-radiation temperatures of 125\,K (red circles) and 11\,K (blue circles). The expected decay corresponding to the fluorescence of the ensemble of initially trapped atoms is displayed as a black line. (b)~Normalized H$^+$ ion signal as a function of PFI delay for an initially excited $n=30$ Rydberg-Stark states measured after cooling the decelerator and trap to  7.5\,K (blue circles), 11.0\,K (black squares) and 21.4\,K (red diamonds).}
\end{figure*}

To illustrate the main features of the trap-decay curves used to study the evolution of the population of Rydberg atoms in the off-axis trap, Fig.~\ref{fig:ch6-temperature}(a) compares the curves measured at room temperature (black circles), $T=125$\,K (red circles), and $T=11$\,K (blue circles) following excitation, deceleration, deflection, and off-axis trapping of low-field-seeking $n=32$, $k=15-23$ Rydberg-Stark states of the hydrogen atom. These temperatures were measured with the sensor labeled T$_6$ in Fig.~\ref{fig:h-setup}(c), and are not the translational temperature ($\approx 150$\,mK) of the trapped atom samples. The full black line in Fig.~\ref{fig:ch6-temperature}(a) represents the exponential decay expected from the radiative (fluorescence to all lower levels allowed by single-photon selection rules) lifetime, which was calculated to be $180\,\mu$s (see Section~\ref{sc:lifetime} A).

As explained in the previous section, the initially prepared Rydberg states can decay by fluorescence, by blackbody-radiation-induced transitions to neighboring $n$ states, by blackbody-radiation-induced photoionization, or as a result of collisions either between a ground-state atom or molecule and the Rydberg atom or between two Rydberg atoms. The observed trap loss results from transitions to states that either do not have an electric dipole moment of the required magnitude and sign or cannot be field ionized by the pulsed electric field.

The trap decay obtained at room temperature is approximately exponential at early times, with a 1/e-time of $135\,\mu$s, significantly shorter than the fluorescence lifetime. Decay processes other than fluorescence therefore play a role in the experiment at room-temperature, even at early times.

When cooling down the environment surrounding the trap from room temperature to $T=11$\,K, the background pressure in the trap, and thus the background ionization signal decreased by more than two orders of magnitude. Consequently, the trap decay curves could be measured for longer times, up to 2\,ms  and $\geq 3$\,ms for the measurements carried out at 125\,K (red circles in Fig.~\ref{fig:ch6-temperature}(a)) and 11\,K (blue circles in Fig.~\ref{fig:ch6-temperature}(a)), respectively.
The initial decay at these lower temperatures is slower than at room-temperature and corresponds closely to the calculated fluorescence rate. However, beyond $\sim 400\,\mu$s, the decay noticeably deviates from an exponential behavior, the effect being more pronounced in the data recorded at 11\,K than at 125\,K.

These results can be qualitatively interpreted as arising from the effects of blackbody radiation on the Rydberg atoms, most importantly blackbody-radiation-induced photoionization. At 300\,K, a substantial fraction of the thermal radiation has a wave number larger than $100\, \text{cm}^{-1}$, which is sufficient to ionize Rydberg atoms with principal quantum number around 30 (the binding energy of an $n=32$ Rydberg electron is $\sim 100\,hc\,\text{cm}^{-1}$). Transitions from the initially prepared $n=32$ Rydberg-Stark states to Rydberg states with $n<23$ can also be stimulated by 300\,K radiation and contribute to the observed decay.
We therefore conclude that stimulated transitions, including blackbody-radiation-induced photoionization, make a significant contribution to the loss of atoms from the trap and to the observed trap-decay rate of ($135\,\mu$s)$^{-1}$.

\begin{figure*}[!tb]
\centering
\includegraphics[angle=270,width=1.0\textwidth]{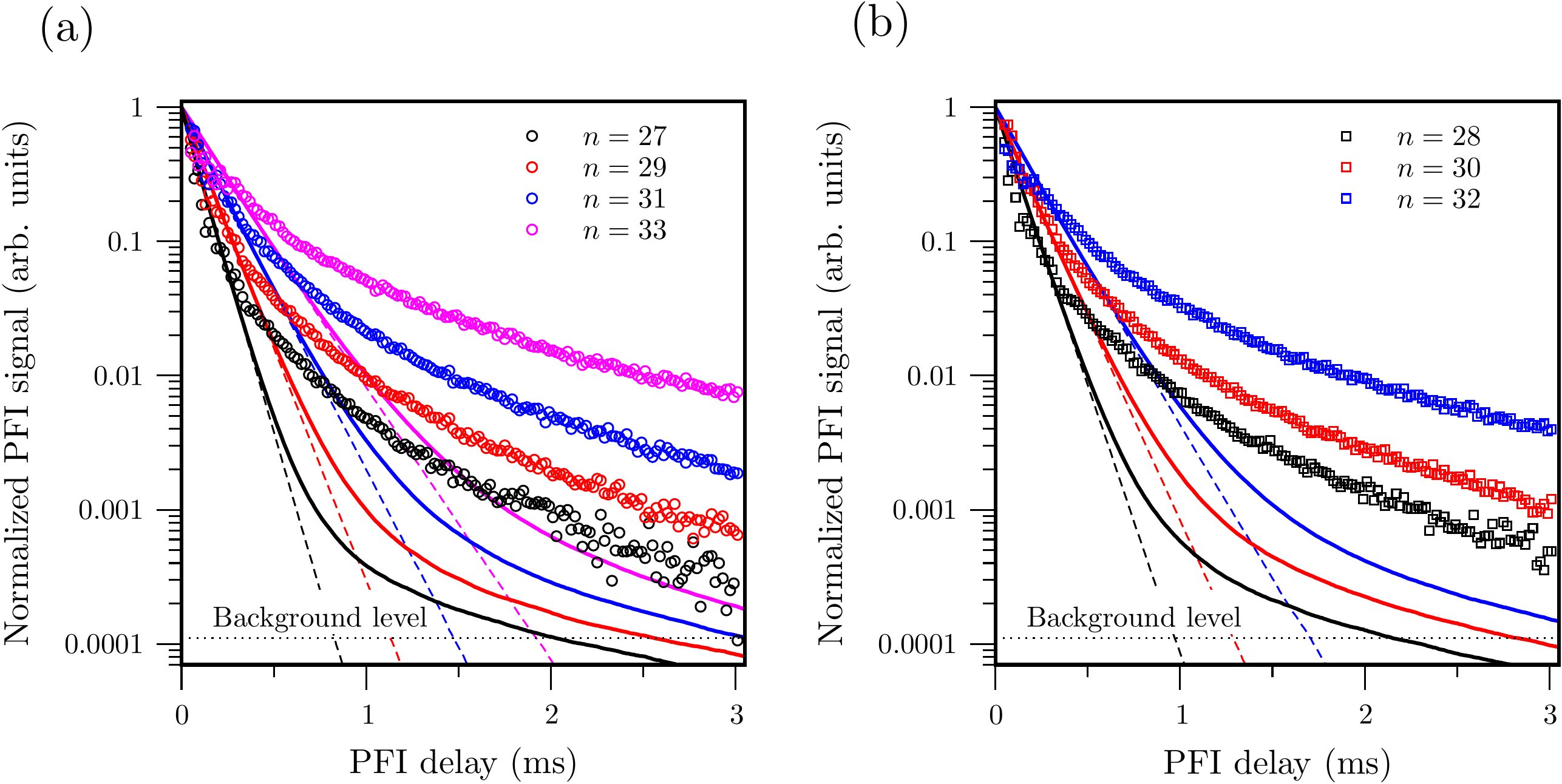}
\caption{\label{fig:ch6-n-dependence-experiment}(a,b)~Loss of hydrogen atoms from the off-axis trap after initial excitation, deceleration, deflection, and trapping of (a)~$n=27$ (black circles), $29$ (red circles), $31$ (blue circles) and $33$ (magenta circles), and (b)~$n=28$ (black squares), $30$ (red squares) and $32$ (blue squares) Rydberg-Stark states at a blackbody-radiation temperature of 11\,K (see text). The trap loss curves, calculated assuming that fluorescence is the only decay process, are indicated by dashed lines. The full lines represent decay curves obtained in calculations including all radiative processes (see Section~\ref{subsc:monte-carlo}).}
\end{figure*}

 At 125\,K, Planck's radiation distribution peaks at a wave number of $\sim 250\,\text{cm}^{-1}$ and the intensity of the blackbody radiation is much lower, so that the importance of blackbody-radiation-induced photoionization and transitions to neighboring Rydberg states is strongly reduced. At 11\,K, photoionization by the blackbody radiation stems exclusively from the 300 K radiation penetrating through the holes of the heat shields (1~\% of the solid angle, see Subsection~\ref{subsc:cryo-setup}) and is not observable.
In both cases, the early decay is now dominated by fluorescence to the ground state or to low Rydberg states that do not have dipole moments large enough for trapping. Transitions to neighboring Rydberg states of similar dipole moments induced by the thermal radiation or by collisions lead, on a longer timescale, to a gradual evolution of the population from the initially prepared Stark states to Stark states of neighboring $n$ values. Because of the rapid increase of the lifetimes with increasing values of $n$ and $|m|$, and of the Rydberg-state spectral density with~$n$, the overall effect is a preferential population of Rydberg states having longer lifetimes, and thus a reduction of the decay rate with time. The observation of transitions between Rydberg states induced by thermal radiation and/or collisions, even at low temperatures, indicates that such transitions must be included in the analysis of the present results and may play a role in other experiments relying on the long lifetimes of Rydberg states, such as H-atom photofragment-translational-spectroscopic experiments~\cite{ashfold11a} and experiments on antihydrogen at CERN~\cite{kellerbauer08a,andresen11a}.

Figure~\ref{fig:ch6-temperature}(b) presents
the trap decay curves of $n=30$ Rydberg-Stark states obtained at temperatures of 7.5\,K (blue circles), 11.0\,K (black squares) and 21.4\,K (red diamonds). Because the intensity of the thermal radiation is strongly suppressed when the temperature is reduced from 21.4\,K to 7.5\,K, one would have expected differences in the trap-decay curves beyond 400 $\mu$s. However, the curves are identical within the precision limits of our measurements, which indicates that the slow redistribution of population observed as deviation from an exponential decay beyond 400 $\mu$s is either caused by the 1\%-solid-angle exposure to room-temperature thermal radiation or by collisions. The following sections will present a quantitative analysis of these results.

The $n$ dependence of the trap decay becomes clearly observable after cooling the device to $T=11$\,K, as illustrated by the measurements presented in Fig.~\ref{fig:ch6-n-dependence-experiment}(a,b). These measurements were carried out following excitation to $n=27$ (black circles), $n=28$ (black squares), $n=29$ (red circles), $n=30$ (red squares), $n=31$ (blue circles), $n=32$ (blue squares) and $n=33$ (magenta circles) Rydberg-Stark states, adjusting the range of $k$ values from 18-26 at $n=27$ to 14-22 at $n=33$, so that the average value of the electric dipole moment was approximately the same in all experiments, i.e., about 900 $a_0e$, as explained in Subsection~\ref{subsc:laser-setup}.
For better visibility, the results obtained for (a)~odd and (b)~even $n$ values are displayed in separate panels. The different trap-decay curves differ both in the rate of the initial exponential decay, which corresponds closely to the expected decay by fluorescence, and in the amplitude of the nonexponential tail beyond 400 $\mu$s. After 3\,ms, the field ionization signal measured in the experiments carried out at $n=33$ is almost two orders of magnitude stronger than at $n=27$. The $n$-dependent behavior of the trap-decay curves presented in Fig.~\ref{fig:ch6-n-dependence-experiment} offers the opportunity to quantitatively analyze and model radiative and collisional processes taking place in the trap on the timescale of several milliseconds.


\section{\label{sc:lifetime} Radiative contributions to the trap decay}

To model the trap-decay curves measured experimentally, it is necessary to consider all possible transitions the atoms undergo, including sequences of transitions, until their dipole moments no longer have the magnitude and sign required for trapping. Some transitions, such as fluorescence to the ground state, immediately lead to trap loss, whereas other transitions only slightly change the dipole moment and do not cause trap loss. To simulate the trap decay, we chose a Monte-Carlo-simulation approach, described in more detail in Subsection~\ref{subsc:blackbody-radiation-tot}, which enabled us to consider the relevant sequences of radiative and collisional processes.

Analytic expressions are available to evaluate the transition electric-dipole moments connecting $\left|n\ell m\right>$ Rydberg states of hydrogen. These expressions can be used to further calculate transition dipole moments between Rydberg-Stark states by using the fact that Stark states, designated here by $\left|n k m\right>$, where $k$ is the difference between the parabolic quantum numbers $n_1$ and $n_2$ ($n_1$ and $n_2$ are related to $n$ and $|m|$ by $n=n_1+n_2+|m|+1$, see Refs.~\cite{bethe57a,gallagher94a}), can be represented as linear combinations of $\left|n\ell m\right>$ basis functions. In this section, we summarize the expressions we used to model all radiative processes, which include spontaneous emission, and absorption and emission stimulated by the thermal radiation field and compare the trap-decay curves calculated using these expressions with the experimental trap-decay curves presented in Section~\ref{sc:results}.

The lifetime of a Rydberg state of H is determined by the spontaneous emission (fluorescence) rates to all lower-lying levels, by transitions to other states stimulated by blackbody radiation, including blackbody-radiation-induced ionization, and by inelastic collisions with other Rydberg atoms or with ground-state atoms or molecules. If the Rydberg states are assumed to be in a collision-free environment, as one might expect, e.g., in a supersonic beam at low Rydberg-atom density, only the radiative processes need to be considered. The total lifetime~$\tau_{\text{tot}}$ of atomic Rydberg states is then determined by the fluorescence (spontaneous emission) lifetime~$\tau_{\text{fl}}$ and by transitions stimulated by the blackbody(BB)-radiation field
\begin{equation}
\frac{1}{\tau_{\text{tot}}} = \frac{1}{\tau_{\text{fl}}}+\frac{1}{\tau_{\text{BB}}^{\Delta n}}+\frac{1}{\tau_{\text{BB}}^{\text{ion}}}~,
\label{eq:tau-tot}
\end{equation}
where the blackbody-radiation-induced transitions have been classified in transitions between bound states ($\tau_{\text{BB}}^{\Delta n}$) and ionizing transitions ($\tau_{\text{BB}}^{\text{ion}}$), a process we refer to as "blackbody-radiation-induced ionization".


\subsection{\label{subsc:spontaneous-emission} Spontaneous emission}

The natural lifetime $\tau_{n\ell m}$ of a Rydberg state~$\left|n\ell m\right>$ can be calculated using
\begin{equation}
\tau_{n\ell m} = \left(\sum_{n'\ell' m'}A_{n'\ell' m',n\ell m}\right)^{-1} \, ,
\label{eq:tau-fluorescence}
\end{equation}
where $A_{n'\ell' m',n\ell m}$ is the Einstein coefficient for the spontaneous emission to an energetically lower-lying state~$\left|n'\ell' m'\right>$. In SI~units, $A_{n'\ell' m',n\ell m}$ is given by~\cite{bethe57a,merkt11b}
\begin{eqnarray}
A_{n'\ell' m',n\ell m} &=& \frac{\omega_{n'\ell' m',n\ell m}^3}{3\hbar c^3\,\pi\epsilon_0}\left|\left<n'\ell' m'\left|\hat{\mu}_{\alpha}\right|n\ell m\right>\right|^2.\nonumber\\
\label{eq:A-spherical}
\end{eqnarray}
In Eq.~(\ref{eq:A-spherical}), $\hat{\mu}_{\alpha}=e\hat{\alpha}$ ($\alpha=x,y,z$) denote the components of the electric-dipole-moment operator
\begin{alignat}{3}
\hat{\mu}_x &= e\hat{x} &\;=\;& er\sin\theta\,\cos\phi\nonumber~,\\
\hat{\mu}_y &= e\hat{y} &\;=\;& er\sin\theta\,\sin\phi\nonumber~,\\
\hat{\mu}_z &= e\hat{z} &\;=\;& er\cos\theta~.
\label{eq:dip-operator-spherical}
\end{alignat}

The Einstein coefficients for spontaneous emission are proportional to the third power of the angular frequency~$\omega_{n'\ell' m',n\ell m}$ corresponding to the energy difference between the two states. Consequently, the decay by fluorescence
strongly favors transitions to the ground state (GS). In the case of H, the transition moments $\left<n'\ell' m'\left|\hat{\mu}_{\alpha}\right|n\ell m\right>$ in Eq.~(\ref{eq:A-spherical}) can be determined analytically as products of a radial and an angular integral \cite{bethe57a,gallagher94a}. In nonhydrogenic atoms, the radial part is determined numerically, e.g., using the Numerov integration method described in Ref.~\cite{zimmerman79a}.
For low-$\ell$ Rydberg states, the scaling with $n$ of $\tau_{n\ell m}$ is primarily determined by the $n^{-3/2}$ dependence of the amplitude of Rydberg-electron wave function in the immediate vicinity of the ion core. Because the transition frequency $\omega_{\text{GS},n\ell m}$ approaches a constant at high-$n$ values, the lifetimes of low-$\ell$ states scale as  \cite{bethe57a,gallagher94a}
\begin{equation}
\tau_{n\ell m} \propto n^3,
\label{eq:tau-fluorescence-scaling}
\end{equation}
which leads to typical radiative lifetimes in the range $1-20\,\mu$s for these states at $n$ values around 30.

The Einstein coefficient~$A_{n'k' m',nk m}$ needed to describe the decay of a Rydberg-Stark state~$\left|nkm\right>$ can be expressed as
\begin{equation}
A_{n'k' m',nkm} = \frac{\omega_{n'k'm',nkm}^3}{3\hbar c^3\,\pi\epsilon_0}
\left|\sum_{\ell=|m|}^{n-1}\sum_{\ell'=|m'|}^{n'-1} \left<n\ell m|nkm\right>\left<n'k'm'|n'\ell' m'\right>\left<n'\ell' m'\left|\hat{\mu}_{\alpha}\right|n\ell m\right>\right|^2~.
\label{eq:A-parabolic}
\end{equation}

For the calculation of the Einstein coefficients of transitions between Stark states, it is convenient to express the dipole operator given in Eq.~(\ref{eq:dip-operator-spherical}) in a form corresponding to linearly ($\hat{\mu}_z$) or circularly ($\hat{\mu}_{x \pm iy}$) polarized radiation
\begin{alignat}{3}
\hat{\mu}_{x+iy} &= \frac{1}{\sqrt{2}}\left(\hat{\mu}_x+\text{i}\hat{\mu}_y\right) &\;=\;& \frac{1}{\sqrt{2}}\,er\sin\theta\,\exp\left\{\text{i}\phi\right\},~\nonumber\\
\hat{\mu}_{x-iy} &= \frac{1}{\sqrt{2}}\left(\hat{\mu}_x-\text{i}\hat{\mu}_y\right) &\;=\;& \frac{1}{\sqrt{2}}\,er\sin\theta\,\exp\left\{-\text{i}\phi\right\}.
\label{eq:dip-operator-parabolic}
\end{alignat}

\begin{figure}[!tb]
\centering
\includegraphics[angle=270,width=0.85\textwidth]{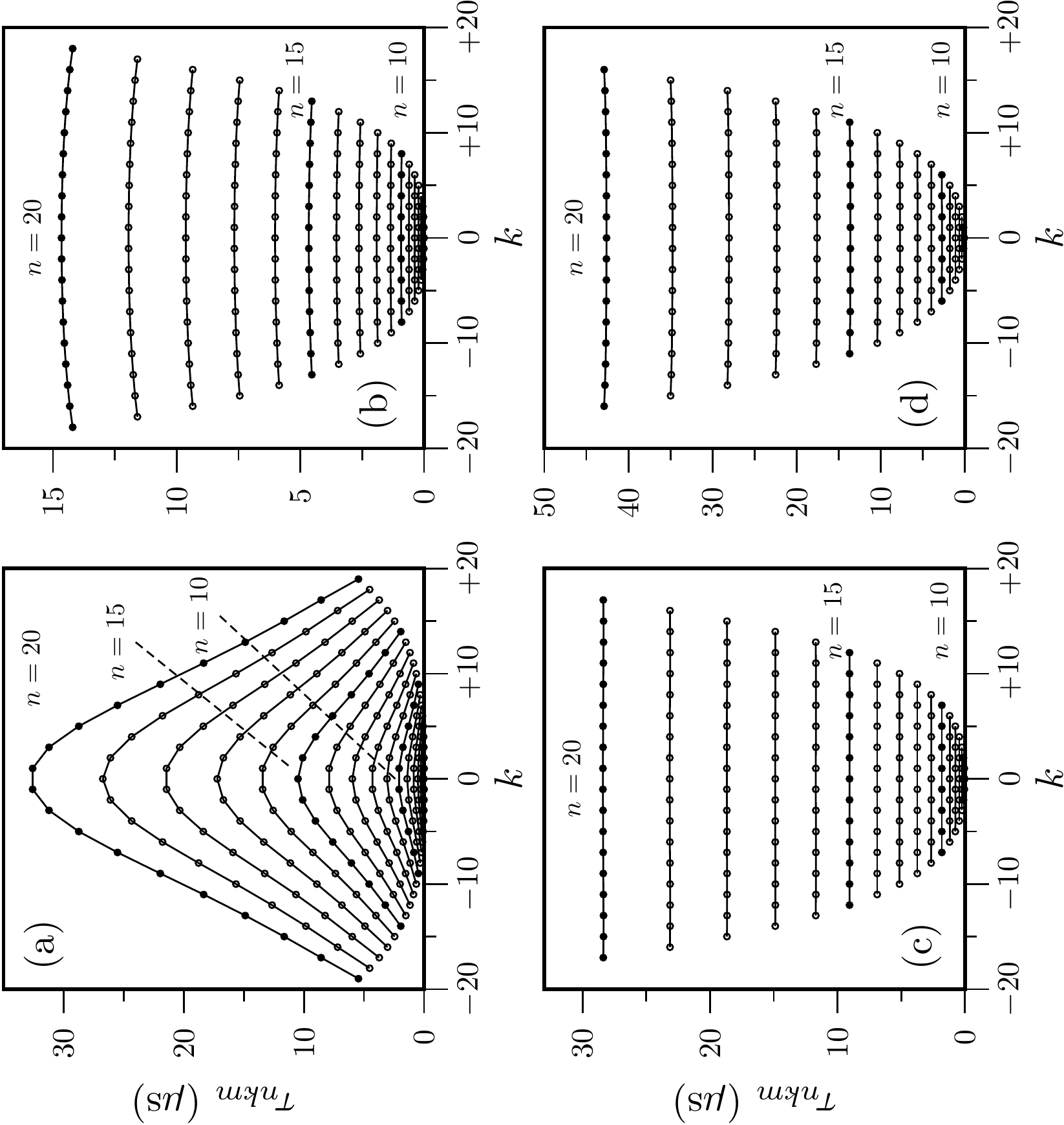}
\caption{\label{fig:fluorescence-f1} Fluorescence lifetimes of $n=2$ to $n=20$ Rydberg-Stark states of H for (a)~$m=0$, (b)~$\left|m\right|=1$, (c)~$\left|m\right|=2$ and (d)~$\left|m\right|=3$.}
\end{figure}

The natural lifetime of a Rydberg-Stark state $\left|nkm\right>$ is given by an equation similar to Eq.~(\ref{eq:tau-fluorescence}) as the inverse sum of all Einstein coefficients for transitions to energetically lower-lying Stark states with quantum numbers $n'$, $k'$ and $m'$
\begin{eqnarray}
\tau_{nkm} &=& \left(\sum_{n'k'm'} A_{n'k'm',nkm}\right)^{-1}.
\label{eq:tau-parabolic}
\end{eqnarray}

Figure~\ref{fig:fluorescence-f1} shows calculated spontaneous-emission lifetimes of Rydberg-Stark states of H with quantum numbers $n$ and $k$ for (a)~$m=0$, (b)~$\left|m\right|=1$, (c)~$\left|m\right|=2$ and (d)~$\left|m\right|=3$.
The contributions to the lifetimes from spontaneous emission to lower Stark states of the same $n$ manifold are negligible at low fields because of the $\omega^3$ scaling of the Einstein $A$ coefficients and are not included.
In the case of $m=0$ states, the lifetimes of states located close to the center of the Stark manifold ($k\approx 0 $) are longer than the lifetimes of states located near the edges of the manifold ($k\approx\pm n$). This behavior originates from the fact that $m=0$ Rydberg-Stark states with $k\approx 0$ have small contributions arising from low-$\ell$ components, especially from $\ell=1$, which is the only component that is connected to the $n=1\, ^2{\rm S}_{1/2}$ ground state by electric-dipole transitions. At higher values of $\left| m\right|$, the $k$ dependence of the fluorescence lifetimes becomes less important: For  $\left| m\right|=1$, the contributions of low-$\ell$ components to the Stark states located near the center of the manifold ($k\approx 0$) are still slightly smaller, and thus the lifetimes slightly longer. For $\left|m\right|\geq 2$, the lifetimes of all Rydberg-Stark states of a given Stark manifold are approximately equal, because the contributions of the $\ell=0$ and $1$ components to the wavefunctions are zero, and the decay to the $^2{\rm S}{1/2}$ electronic ground state of atomic hydrogen is suppressed.

Figure~\ref{fig:fluorescence-f1} also illustrates that the lifetimes of Rydberg-Stark states with $\left|m\right| \le 3$ rapidly
increase with increasing $n$ values and scale as $n^4$. The additional factor of $n$ compared to Eq.~(\ref{eq:tau-fluorescence-scaling}) results from the dilution of the low-$\ell$ components of the wave function as the number of high-$\ell$ components of the Stark state increases ($\ell_{\rm max}=n-1$). Typical spontaneous-emission lifetimes of Rydberg-Stark states with $n=20$ are $\approx 15\,\mu$s for $m=0$, $\approx 15\,\mu$s for $\left|m\right|=1$, $\approx 30\,\mu$s for $\left|m\right|=2$, and $\approx 40\,\mu$s for $\left|m\right|=3$. At $n=30$, these lifetimes are already $\approx 70\,\mu$s for $m=0$, $\approx 70\,\mu$s for $\left|m\right|=1$, $\approx 140\,\mu$s for $\left|m\right|=2$, and $\approx 210\,\mu$s for $\left|m\right|=3$.

\begin{figure}[!tbh]
\centering
\includegraphics[angle=270,width=0.85\textwidth]{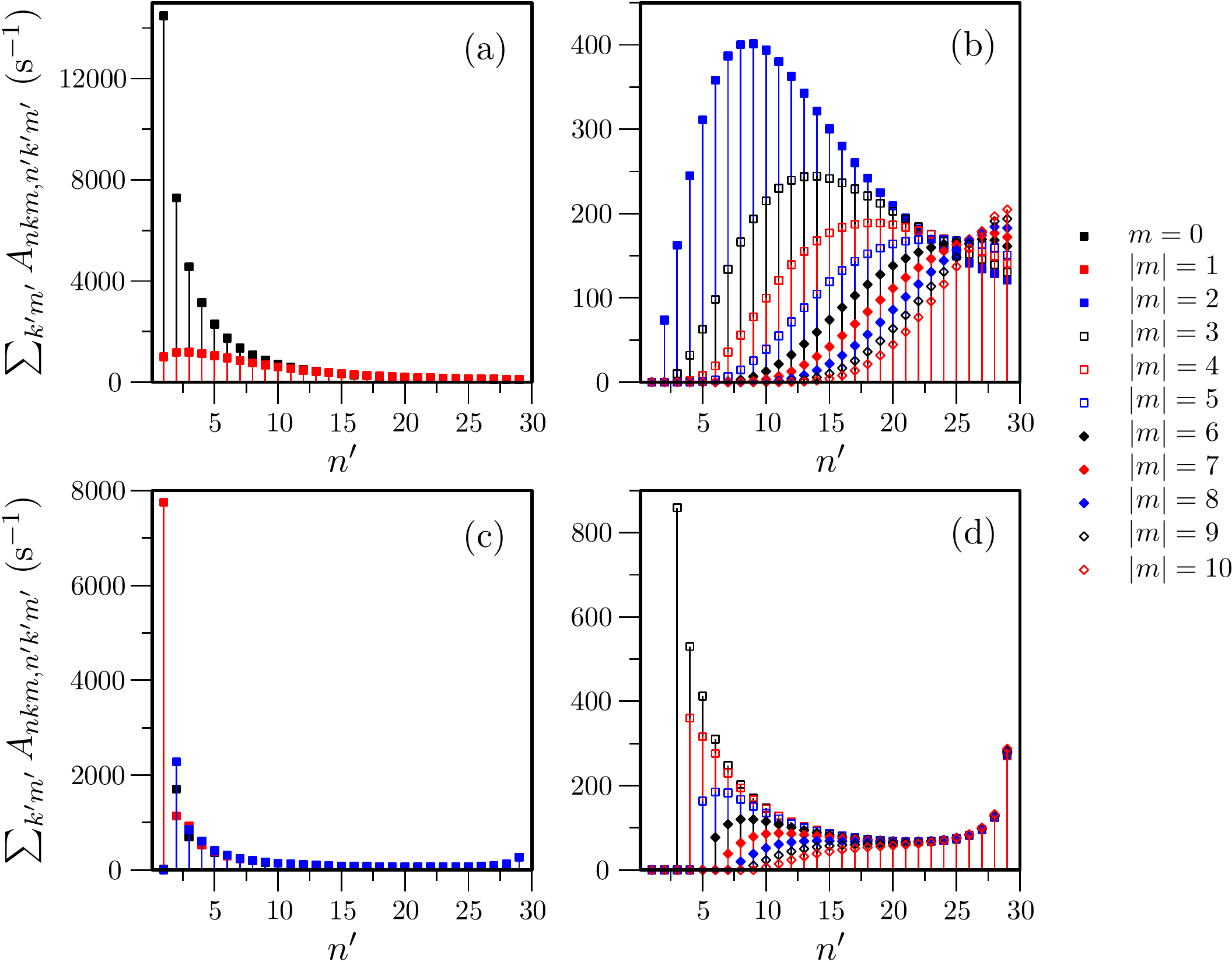}
\caption{\label{fig:fluorescence-f2} Sum $\sum_{k'm'} A_{nkm,n'k'm'}$ of the spontaneous-emission rates from selected $k$ and $m$ Stark states of the $n=30$ manifold to all optically accessible Rydberg-Stark states of principal quantum number $n'$. Panels (a) and (b) display the behavior of the outermost Stark states and panels (c) and (d) displays the behavior of the central Stark states. Note that very different vertical scales have been used for the lowest $\left|m\right|$ states (panels (a) and (c)) and for the other $\left|m\right|$ states (panels (b) and (d)).}
\end{figure}
\begin{figure*}[!tbh]
\centering
\includegraphics[angle=270,width=1.0\textwidth]{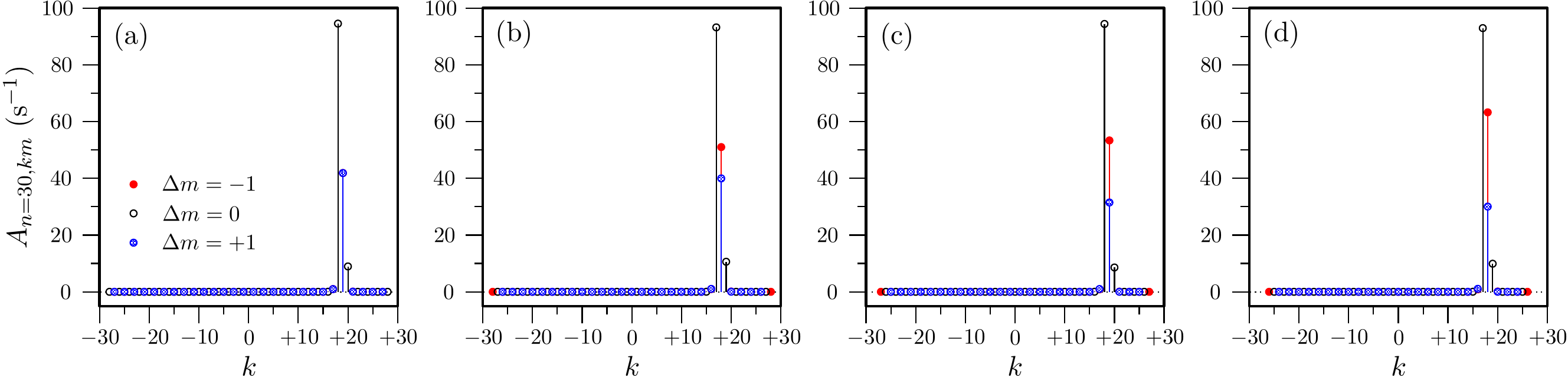}
\caption{\label{fig:fluorescence-f3} Fluorescence rates corresponding to (a)~$\left|n=30,k=19,m=0\right>\rightarrow\left|n'=29,k',m'\right>$, (b)~$\left|n=30,k=18,\left|m\right|=1\right>\rightarrow\left|n'=29,k',m'\right>$, (c)~$\left|n=30,k=19,\left|m\right|=2\right>\rightarrow\left|n'=29,k',m'\right>$, and (d)~$\left|n=30,k=18,\left|m\right|=3\right>\rightarrow\left|n'=29,k',m'\right>$ transitions. Transitions which induce a change in $\left|m\right|$ are displayed separately for $\Delta m=-1$ (red), $\Delta m=0$ (black) and $\Delta m=+1$ (blue).}
\end{figure*}

 To illustrate the dependence of the distribution of principal quantum numbers $n'$ resulting from spontaneous emission from selected Rydberg-Stark states, Fig.~\ref{fig:fluorescence-f2} displays the sum $\sum_{k'm'} A_{n'k'm',nkm}$ of Einstein $A$ coefficients for the most central Stark states ($k=0$ or $\pm 1$, depending on the value of $\left|m\right|$, panels (a) and (b)) and for the extreme Stark states ($k_{\rm max}= n-\left|m\right| -1$, panels (c) and (d)) of the $n=30$ Stark manifold.
 The transitions between Stark states of the $n=30$ manifold are not considered, because their $A$ coefficients are strongly field dependent and their $\omega^3$ scaling implies negligible values for the electric fields of less than 50~V/cm relevant to our experiments.
 The Einstein $A$ coefficients decrease very rapidly with $\left|m\right|$ at low $\left|m\right|$ values and the results obtained for the lowest $\left|m\right|$ values are presented on a strongly compressed vertical scale (panels (a) and (c)). At the lowest $\left|m\right|$ values, spontaneous emission strongly favors the population of low $n'$ states because the values of the Einstein coefficients are dominated by their $\omega^3$ dependence (see Eq.~\ref{eq:A-parabolic}). At higher $\left|m\right|$ values, the emission to the lowest $n'$ states is forbidden so that the transition frequencies are strongly reduced and the values of the Einstein $A$ coefficients start being governed by the $n^2$ dependence of the transition moments between Rydberg states of neighboring $n$ values. Figure~\ref{fig:fluorescence-f2} thus nicely illustrates the evolution of the spontaneous decay rates from the transition-frequency-dominated regime at low $\left|m\right|$ values, which favors transitions with the largest possible changes of $\Delta n=n - n'$, to the transition-moment-dominated regime at high $\left|m\right|$ values, which favors transitions with the smallest possible changes of $\Delta n$.
 Fig.~\ref{fig:fluorescence-f3} demonstrates that transitions associated with small $\Delta n$ values do not significantly change $k$ and thus preserve the electric dipole moment. Consequently, a single such transition does not cause a loss of Rydberg atoms from the trap.

 One may therefore conclude that rapid trap loss by spontaneous emission primarily affects Rydberg-Stark states with $\left|m\right| \leq 1$. Spontaneous emission from states with $\left|m\right| \geq 2$, on the other hand, can contribute to a gradual evolution of the initially prepared Rydberg-state population to states of higher $\left|m\right|$ values but similar dipole moments, which are longer lived and remain trapped for longer times. $\left|m\right|$-changing but dipole-moment-preserving spontaneous emission therefore represents a mechanism by which a slowly decaying nonexponential component in the trap-loss curves as observed experimentally in Figs.~\ref{fig:ch6-temperature} and  \ref{fig:ch6-n-dependence-experiment} might be explained.


\subsection{\label{subsc:blackbody-radiation} Transitions stimulated by thermal radiation}

The spacing between two adjacent Rydberg states scales as $2R/n^3$ and is small at high $n$ values, e.g., $\approx 8\,\text{cm}^{-1}$ at $n=30$. Consequently, the thermal radiation from the environment can induce transitions between Rydberg-Stark states, even if the experiments are carried out under cryogenic conditions. If the energy of a photon from the blackbody radiation field exceeds the binding energy of the Rydberg electron ($\approx (hc)120\,\text{cm}^{-1}$ at $n=30$), absorption can result in the ionization of the Rydberg atom and must also be considered in the analysis of the measured trap-decay curves.  Similarly, photons from the thermal radiation field with wave numbers sufficiently high to stimulate emission to Rydberg states with $n<23$ can also contribute to the observed trap decay.

The rate of absorption or stimulated emission of an atom in a specific Rydberg-Stark state $\left|n\ell m\right>$ to other Stark states $\left|n'\ell' m'\right>$ is conveniently described by a matrix ${\bf K}$ with elements
\begin{eqnarray}
K_{n'k'm',nkm} &=& \overline{n}\left(\omega_{n'k'm',nkm}\right)\,A_{n'k'm',nkm},\nonumber\\
\label{eq:Einstein-B}
\end{eqnarray}
where $\overline{n}(\omega_{n'k'm',nkm})$ is the average number of photons at the transition angular frequency $\omega_{n'k'm',nkm}$ which, for thermal radiation, can be described by
\begin{eqnarray}
\overline{n}\left(\omega_{n'k'm',nkm}\right) = \frac{1}{\exp\left\{\frac{\hbar\omega_{n'k'm',nkm}}{k_{\text{B}}T}\right\}-1},\nonumber \\
\label{eq:photon-occupation-number}
\end{eqnarray}
where $k_{\text{B}}$ is Boltzmann's constant and $T$ is the temperature of the thermal radiation field. The stimulated absorption or emission rates thus follow the general trends for $A_{n'k'm',nkm}$ discussed in the previous subsection, but are now weighted by the average photon number at the relevant transition frequency.

\begin{figure}[!tbh]
\centering
\includegraphics[angle=270,width=0.70\textwidth]{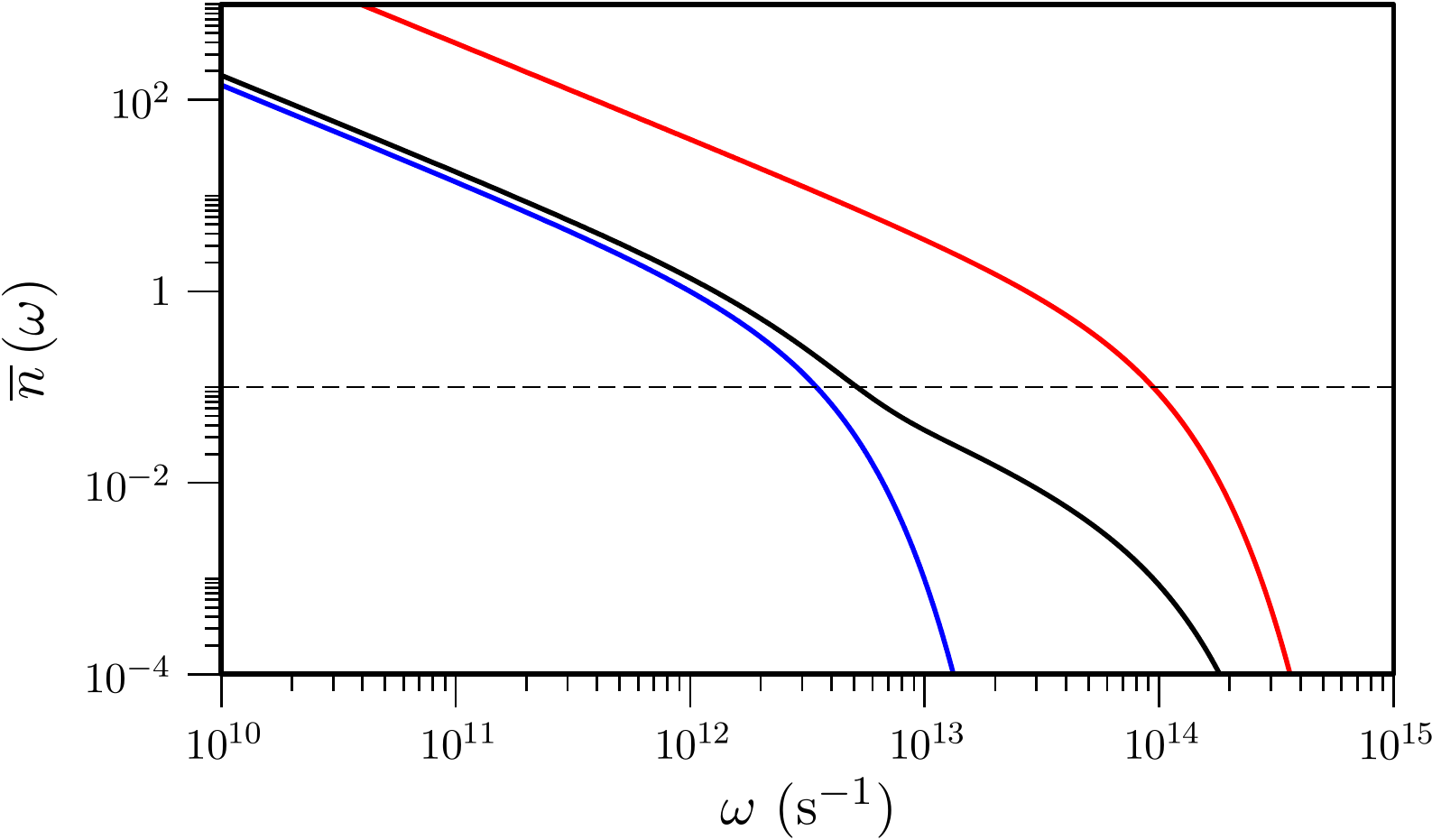}
\caption{\label{fig:photon-occupation-number} Average photon number as a function of the angular frequency at $T=300$\,K (red), $T=11$\,K (blue) and for a blackbody-radiation field consisting of a 1\% contribution at $T=300$\,K and a 99\% contribution at $T=11$\,K (black).}
\end{figure}

The average number of photons strongly depends on the frequency and on the temperature, as illustrated in Fig.~\ref{fig:photon-occupation-number}, which compares the values obtained at $T=300$~K (red trace) at 11~K (blue trace), and for the radiation field corresponding to 11~K with a 1 \% contribution of 300~K radiation relevant for our experiments when the region surrounding the Rydberg-atom trap is cooled to 11~K (see Section~\ref{subsc:cryo-setup}). When the temperature of the radiation field is reduced, the maximum of the Planck distribution rapidly shifts toward lower frequencies until the corresponding energies are not sufficient to ionize the Rydberg states, but can only induce transitions between Rydberg-Stark states of neighboring $n$ values. At low average photon numbers, stimulated emission becomes negligible compared to spontaneous emission and absorption processes become unlikely.

Considering transitions at frequencies corresponding to low (i.e., 0.1, indicated by a dashed horizontal line in Fig.~\ref{fig:photon-occupation-number} or less) average numbers of photons as unlikely, it can be deduced from Fig.~\ref{fig:photon-occupation-number} that the radiation field at 11\,K can only efficiently drive transitions up to frequencies of $\approx 3\cdot 10^{12}\,\text{s}^{-1}$ (i.e. $\approx 16\,\text{cm}^{-1}$), which at $n\approx 30$ primarily leads to small changes $\Delta n=n'-n$ of the principal quantum number. At 300\,K, the radiation field can drive transitions up to angular frequencies of $\approx  10^{14}\,\text{s}^{-1}$ (i.e. $\approx 530\,\text{cm}^{-1}$), which is sufficient to ionize the Rydberg  atoms in the trap or to stimulate transitions from $n=30$ Rydberg states to states with $n<23$.

\begin{figure}[!tbh]
\centering
\includegraphics[angle=270,width=0.75\textwidth]{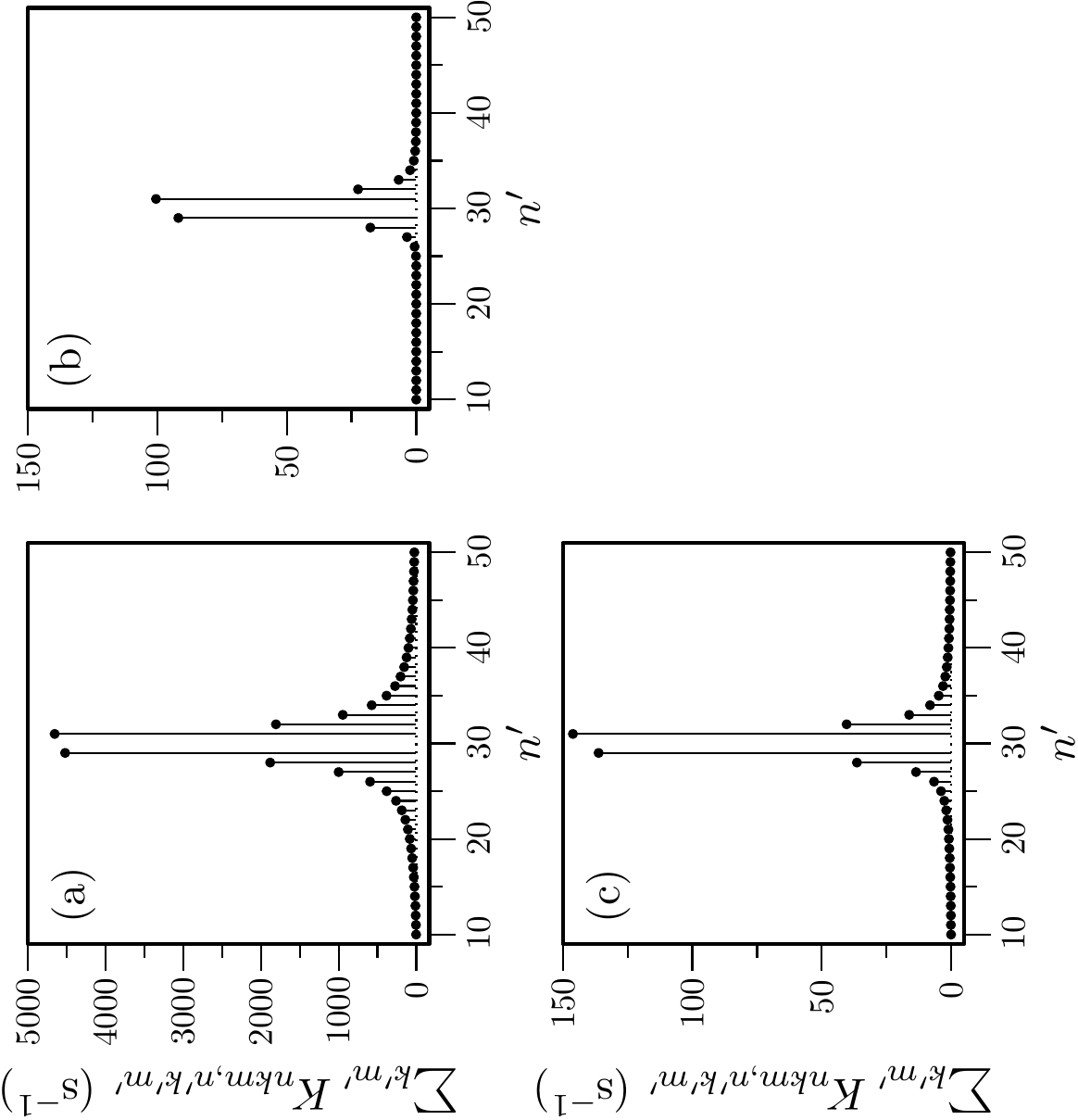}
\caption{\label{fig:blackbody_f1} Sum $\sum_{k'm'} K_{nkm,n'k'm'}$ of the rates of the
blackbody-radiation-induced transitions from the $\left|n=30,k=19,m=0\right>$ Rydberg-Stark state to all accessible $k'm'$ Stark states of a given $n'$ value for temperatures of (a)~$T=300$\,K, (b)~$T=11$\,K, and (c) for a radiation field consisting of a dominant (99\%) $T=11$\,K contribution and a weak (1\%) $T=300$\,K contribution, corresponding to the conditions under which the data in Fig.~\ref{fig:ch6-n-dependence-experiment} were recorded.}
\end{figure}

\begin{figure}[!tbh]
\centering
\includegraphics[angle=270,width=0.75\textwidth]{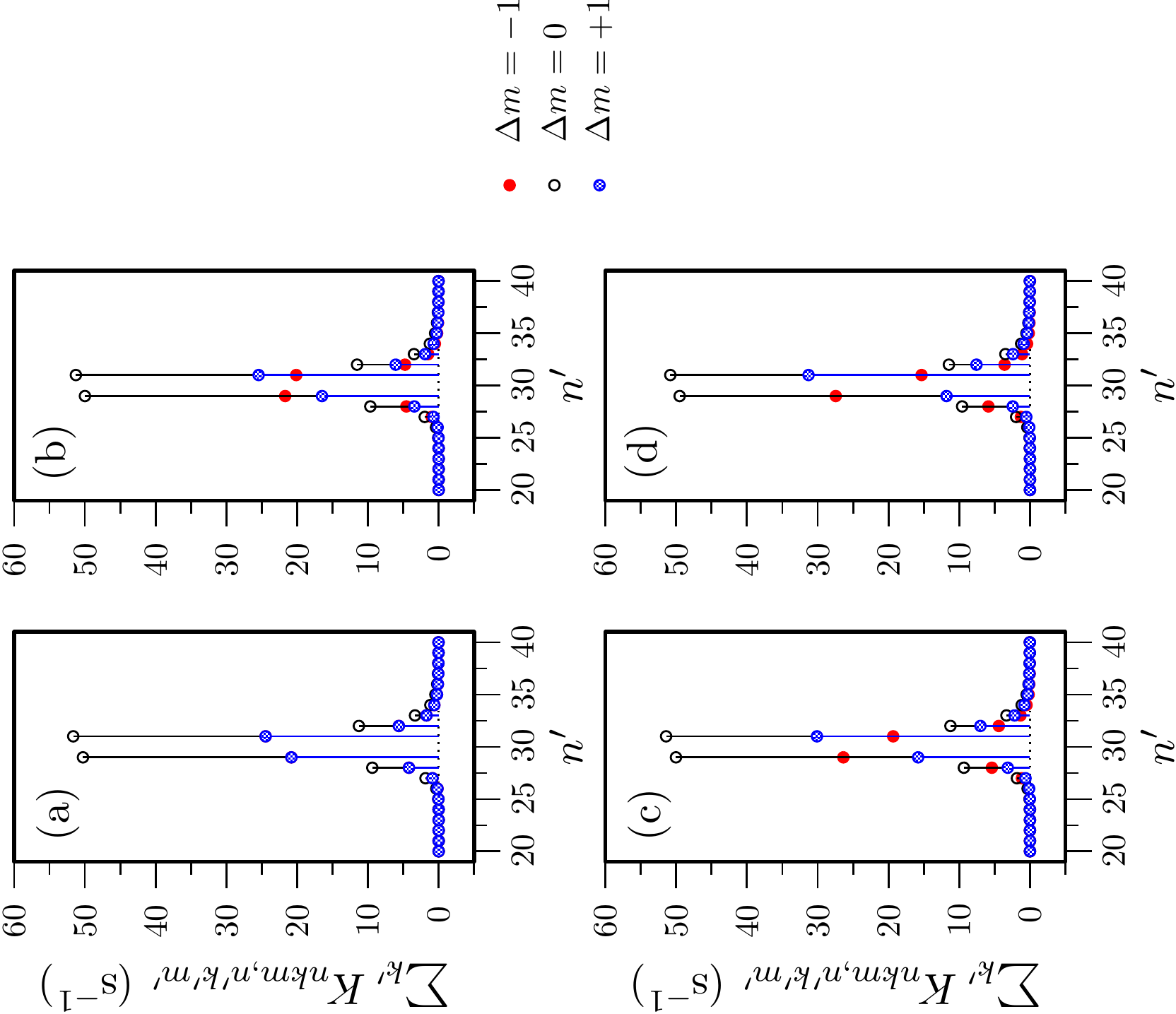}
\caption{\label{fig:blackbody_f2}
Sum $\sum_{k'} K_{nkm,n'k'm'}$ of the rates of the
blackbody-radiation-induced transitions
from the (a)~$\left|n=30,k=19,m=0\right>$, (b) $\left|n=30,k=18,\left|m\right|=1\right>$, (c) $\left|n=30,k=19,\left|m\right|=2\right>$, and (d) $\left|n=30,k=18,\left|m\right|=3\right>$ Rydberg-Stark states at $T=11$\,K. Transitions which induce a change in $\left|m\right|$ are displayed separately for $\Delta m=-1$ (red), $\Delta m=0$ (black) and $\Delta m=+1$ (blue).}
\end{figure}

Figure~\ref{fig:blackbody_f1} displays the sum $\sum_{k'm'} K_{n'k'm',nkm}$ of the rates calculated for stimulated transitions from the $\left|n=30,k=19,m=0\right>$ Rydberg-Stark state to all accessible $k'm'$ levels of selected $n'$ value corresponding to the experimental conditions used to record the data presented in Section~\ref{sc:results}.
At $T=300$\,K, the overall rate rapidly decreases from about 4500\,s$^{-1}$ for $\Delta n=\pm 1$ processes to less than 100\,s$^{-1}$ for $\Delta n=\pm 10$. At 11\,K, the rates are almost two orders of magnitude smaller and transitions with $\left|\Delta n\right| \ge 4$ are entirely suppressed on the timescale of up to 3\,ms over which our trap-decay measurements could be carried out. For the  radiation field consisting of a dominant (99\%) $T=11$\,K contribution and a weak (1\%) $T=300$\,K contribution, corresponding to the conditions under which the data in Fig.~\ref{fig:ch6-n-dependence-experiment} were recorded, the overall rates decrease from  about 150\,s$^{-1}$ for $\Delta n = \pm 1$ to less than 10\,s$^{-1}$ for $\left|\Delta n\right| \ge 4$. Moreover, the rate of stimulated emission between Stark states of the same $n$ manifold is negligible at low fields. For instance, at $n=30$ and for a field of 20 V/cm, $\Delta n = 0, \Delta k= \pm 1$ transitions have wave numbers of about 0.03~cm$^{-1}$ for which $\overline{n} \approx 10^3$ at 11~K. For a $\Delta n =  \pm 1$ transition, the wave number and $\overline{n}$  are approximately 8~cm$^{-1}$ and 1, respectively, so that the former transitions are more than 100 times less likely than the latter.
These considerations enable one to considerably reduce the number of transitions that need to be considered when modeling the contribution of stimulated transitions to the trap-decay.

Stimulated transitions can change the value of $\left|m\right|$ and thereby strongly affect the trap-decay dynamics. Indeed, the spontaneous decay to low-$n$ states critically depends on the value of $\left|m\right|$ (see Fig.~\ref{fig:fluorescence-f2}), so that a stimulated absorption or emission process associated with an increase (decrease) of $\left|m\right|$ can inhibit (induce) the loss of an atom from the trap.
Figure~\ref{fig:blackbody_f2} illustrates the relative importance of $\Delta\left|m\right|=0,\pm 1$ to the overall rates displayed in Fig.~\ref{fig:blackbody_f1}(b) for selected Stark states at $T=11$\,K. In all cases, $\Delta m=0$ transitions (black) are the most probable, and account for about half of the overall rate. $\Delta\left|m\right|=-1$ transitions are more probable than $\Delta\left|m\right|=+1$ for stimulated emission processes and the opposite is true for absorption.  Consequently, stimulated emission to Rydberg-Stark states of neighboring $n$ value do not only facilitate a subsequent spontaneous decay by reducing $n$ but also by reducing $\left|m\right|$, whereas absorption of low-frequency radiation inhibits the spontaneous decay by increasing both $n$ and $\left|m\right|$.

Absorption can also ionize the Rydberg atoms.
At a given temperature of the radiation field, the relative contribution of ionization to the total rate of all stimulated transitions is approximately constant over the range of $n$ values of interest ($n=23-34$) and very rapidly decreases with decreasing temperature. For $n$d states of H, relative ionization contributions of $\approx 9$\% and $\approx 2$\% are predicted at 300\,K and 100\,K, respectively~\cite{glukhov10a}.

\subsection{\label{subsc:blackbody-radiation-tot} Overall radiative decay rates and Monte-Carlo simulation of radiative processes in the trap}

Combining the results of the previous two subsections, the overall decay rate $K_{nkm}^{\text{tot}}$ of a given $\left|nkm\right>$  Rydberg-Stark state can be described by the expression
\begin{equation}
K_{nkm}^{\text{tot}} = \sum_{n'k'm'}A_{nkm,n'k'm'} + \left(1+R_{\text{ion}}\left(n,T\right)\right)\sum_{n'k'm'}K_{nkm,n'k'm'}.
\label{eq:total-decay}
\end{equation}
In this equation, blackbody-radiation-induced ionization is accounted for by introducing the factor $R_{\text{ion}}(n,T)$ in the second term on the right-hand side. Over the range of principal quantum number relevant for the experiments described in Section~\ref{sc:results}, the rate of ionization by the thermal radiation is to a good approximation independent of $n$ so that $R_{\text{ion}}(n,T)$ reduces to a temperature-dependent factor. For the temperatures of 300\,K, 125\,K and 11\,K studied experimentally, its value can be estimated from the calculations for low-$\ell$ states presented in Ref.~\cite{glukhov10a}  to be approximately 0.1, 0.03 and 0, respectively. Consequently, ionization by the thermal radiation plays no role in the experiments carried out at 11\,K, as already pointed out in the discussion of the results presented in Fig.~\ref{fig:ch6-n-dependence-experiment}.

\begin{figure*}[!h]
\centering
\includegraphics[angle=0,width=1.0\textwidth]{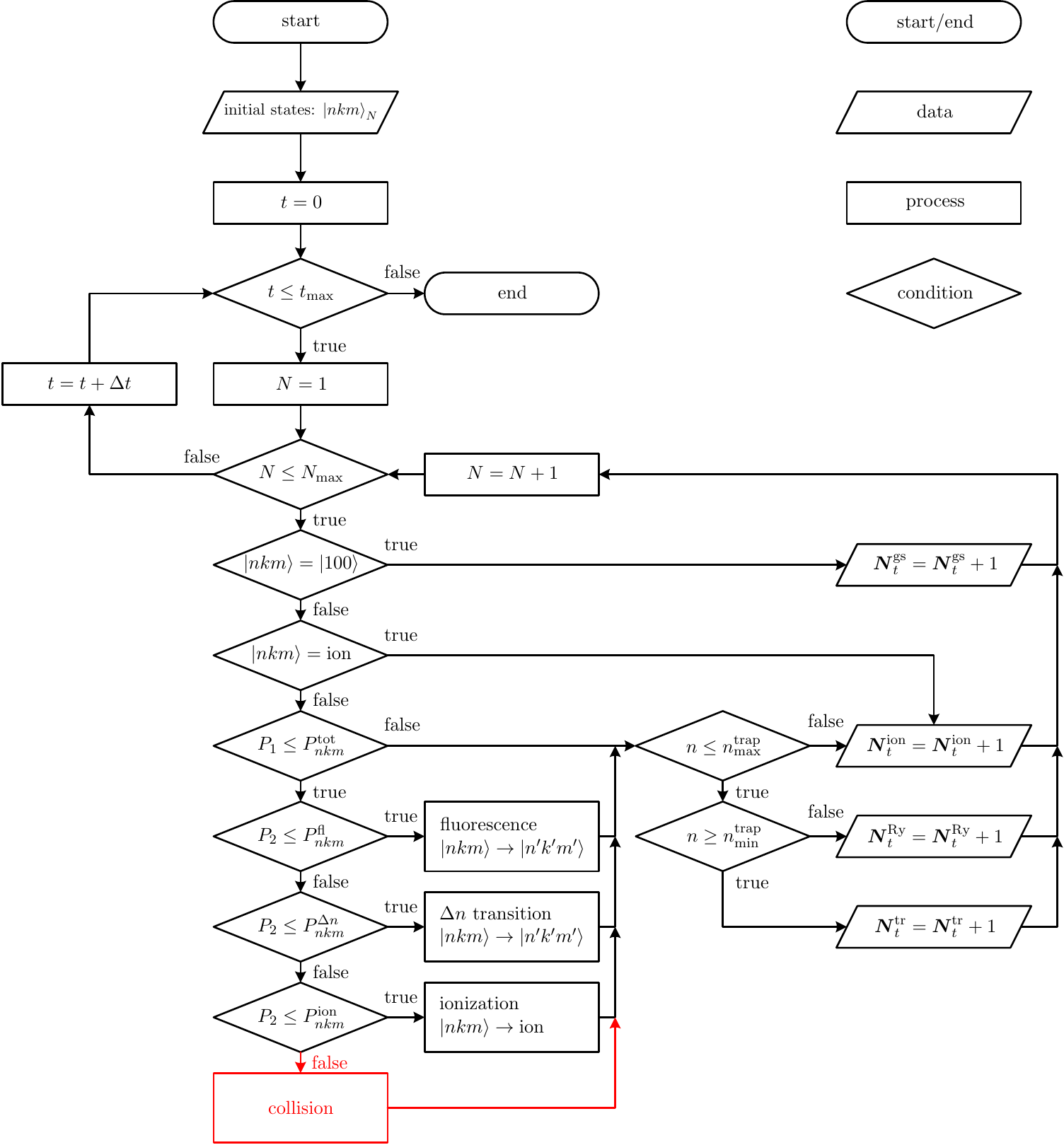}
\caption{\label{fig:flow-diagram} Flow diagram of the Monte-Carlo simulation program used to calculate the decay of trapped hydrogen atoms in Rydberg-Stark states. $N_t^{\text{tr}}$, $N_t^{\text{Ry}}$, $N_t^{\text{gs}}$ and $N_t^{\text{ion}}$ represent the number of trapped Rydberg atoms, the number of Rydberg atoms not fulfilling the trapping conditions, and the number of ions produced by blackbody-radiation-induced ionization, respectively (see text for details). The procedure followed to evaluate the effects of collisions between trapped Rydberg atoms (routines marked in red) is described in Section~\ref{sec:ch6-theory}.}
\end{figure*}
The parallel nature of the decay processes and the form of Eq.~(\ref{eq:total-decay}) are well suited for a Monte-Carlo simulation of the radiative processes taking place in the trap. It also enables one to include the radiative processes by which a given Rydberg-Stark state is populated by a radiative transition in a straightforward manner. The procedure we implemented is illustrated as flow diagram in Fig.~\ref{fig:flow-diagram}. For each particle (index $N$ running from 1 to $N_{\rm max}$), the initial state $\left|nkm\right>$ is chosen to correspond to the Rydberg-Stark states populated by the laser excitation sequence, i.e., to states with a well-defined $n$ value and a narrow range of five to six $k$ values  precisely determined from the known electric field strength in the photoexcitation region, the known structure of Rydberg-Stark states of atomic hydrogen, and the known laser bandwidth (see Sections~\ref{sc:experiment}
and~\ref{sc:results} for details). Because both photons in the photoexcitation sequence are polarized perpendicularly to the electric-field vector at excitation, each induces a $\Delta m=\pm 1$ transition, leading to the population of $\left|nkm=0, \pm2\right>$ Rydberg-Stark states with  the probabilities of 0.18 ($m=0$) and 0.82 ($\left|m\right|=2$).

In the first step of the simulation, a randomly chosen weighted probability $P_1$ is compared to the probability
\begin{equation}
P_{nkm}^{\text{tot}} = K_{nkm}^{\text{tot}}\,\Delta t
\end{equation}
that the initially prepared Stark state decays during the time interval $\Delta t$, which is chosen sufficiently small to ensure convergence of the calculations.
If $P_1>P_{nkm}^{\text{tot}}$, the particle remains in the same state. If $P_1\leq P_{nkm}^{\text{tot}}$, a decay is assumed to occur, which is specified by comparing a second random probability~$P_2$ with weighted probabilities for the three possible decay processes
\begin{align}
P_{nkm}^{\text{fl}}	&= \frac{\sum_{n'k'm'}A_{nkm,n'k'm'}}{K_{nkm}^{\text{tot}}}, \label{14} \\
P_{nkm}^{\Delta n}	&= \frac{\sum_{n'k'm'}A_{nkm,n'k'm'} + \sum_{n'k'm'}K_{nkm,n'k'm'}}{K_{nkm}^{\text{tot}}},\label{15} \\
P_{nkm}^{\text{ion}}	&= 1-P_{nkm}^{\text{fl}}-P_{nkm}^{\Delta n}. \label{16}
\end{align}

If fluorescence takes place ($P_2\leq P_{nkm}^{\text{fl}}$), the final state $\left|n'k'm'\right>$ is determined using a further random probability (not depicted in Fig.~\ref{fig:flow-diagram})
\begin{equation}
P_2' \leq \frac{R_2'}{\sum_{n'k'm'}A_{nkm,n'k'm'}},
\label{eq:p2prime}
\end{equation}
which is evaluated in a loop with index $i$. The sum $R_2'= {\sum_{i=n'k'm'}A_{nkm,n'k'm'}}$ is calculated over all lower-lying electronic states from $\left|n'=1,k'=0,m'=0\right>$ to $\left|n'=n,k'=k,m'=m\right>$ in order of increasing energy. The final state~$\left|n'k'm'\right>$ corresponds to the first Rydberg-Stark state for which the inequality in Eq.~(\ref{eq:p2prime}) is not fulfilled. Most atoms either fluoresce to the $n=1$ ground state or to a very low $n$ state and only a very small fraction of the atoms (typically $<0.01\%$) ends up, after spontaneous emission, in Rydberg-Stark states that fulfill the trapping conditions.

If the particle does not decay by fluorescence ($P_2>P_{nkm}^{\text{fl}}$), a blackbody-radiation-induced process occurs. If $P_2\leq P_{nkm}^{\Delta n}$ (see Eq.~(\ref{15})) a transition to a bound level (superscript $\Delta n$) occurs, otherwise ionization takes place (see Eq.~(\ref{16})). In the latter case, the atom is lost from the trap. In the former case, the final state $\left|n'k'm'\right>$ of the transition is determined with the aid of another random probability $P_3'$ (not depicted in Fig.~\ref{fig:flow-diagram})
\begin{equation}
P_3' \leq \frac{R_3'}{\sum_{n'k'm'}K_{nkm,n'k'm'}},
\label{eq:p3prime}
\end{equation}
following the same procedure as  for the evaluation of Eq.~(\ref{eq:p2prime}). Eq.~(\ref{eq:p3prime}) is evaluated in a loop with index~$j$ and $R_3'=\sum_{j=n'k'm'} K_{nkm,n'k'm'}$ is a sum over all accessible states from $\left|n'=1,k'=0,m'=0\right>$ to $\left|n'=n_{\text{max}},k',m'=m_{\text{max}}\right>$. The final state~$\left|n'k'm'\right>$ corresponds to the first Rydberg-Stark state for which Eq.~(\ref{eq:p3prime}) is not fulfilled, and can be either a lower-lying ($n'<n$, stimulated emission) or a higher-lying ($n'>n$, absorption) Rydberg-Stark state. For the Rydberg-Stark states in the range $n=25-35$ studied experimentally, it turned out to be sufficient to only consider states with $n \leq 70$ and $\left|m\right| \leq 15$ in the simulations.

In the cases where the radiative processes populate another Rydberg-Stark state $\left|n'k'm'\right>$, one must
evaluate whether this state fulfills the trapping conditions or whether it is field-ionized. It turns out, for our experimental conditions, to be sufficient to test whether $n'$ lies in the range 23-65.
Indeed, the values of $k$ and thus the values of the electric dipole moments are not significantly altered by radiative processes connecting neighboring Rydberg states (see Fig.~\ref{fig:fluorescence-f3}).
The simulation program runs through all $N_{\text{max}}$ particles, stores their final state at each time step, updates the number of atoms that are ionized ($N_t^{\rm ion}$), have returned to the 1~$^2{\rm S}{1/2}$ ground state ($N_t^{\rm gs}$), are still in a Rydberg state fulfilling ($N_t^{\rm tr}$) or not fulfilling ($N_t^{\rm Ry}$) the trapping condition, and moves to the next time step. The procedure is repeated until the final time $t_{\text{max}}$ is reached. The status "not fulfilling the trapping condition" is a transient status, because radiative transitions can either enable the particles to be retrapped (which is very unlikely) or to decay to the ground state.

\subsection{\label{subsc:monte-carlo} Quantitative predictions and analysis of the role of radiative processes in Rydberg-atom trapping experiments}

\begin{figure}[!tbh]
\centering
\includegraphics[angle=270,width=0.5\textwidth]{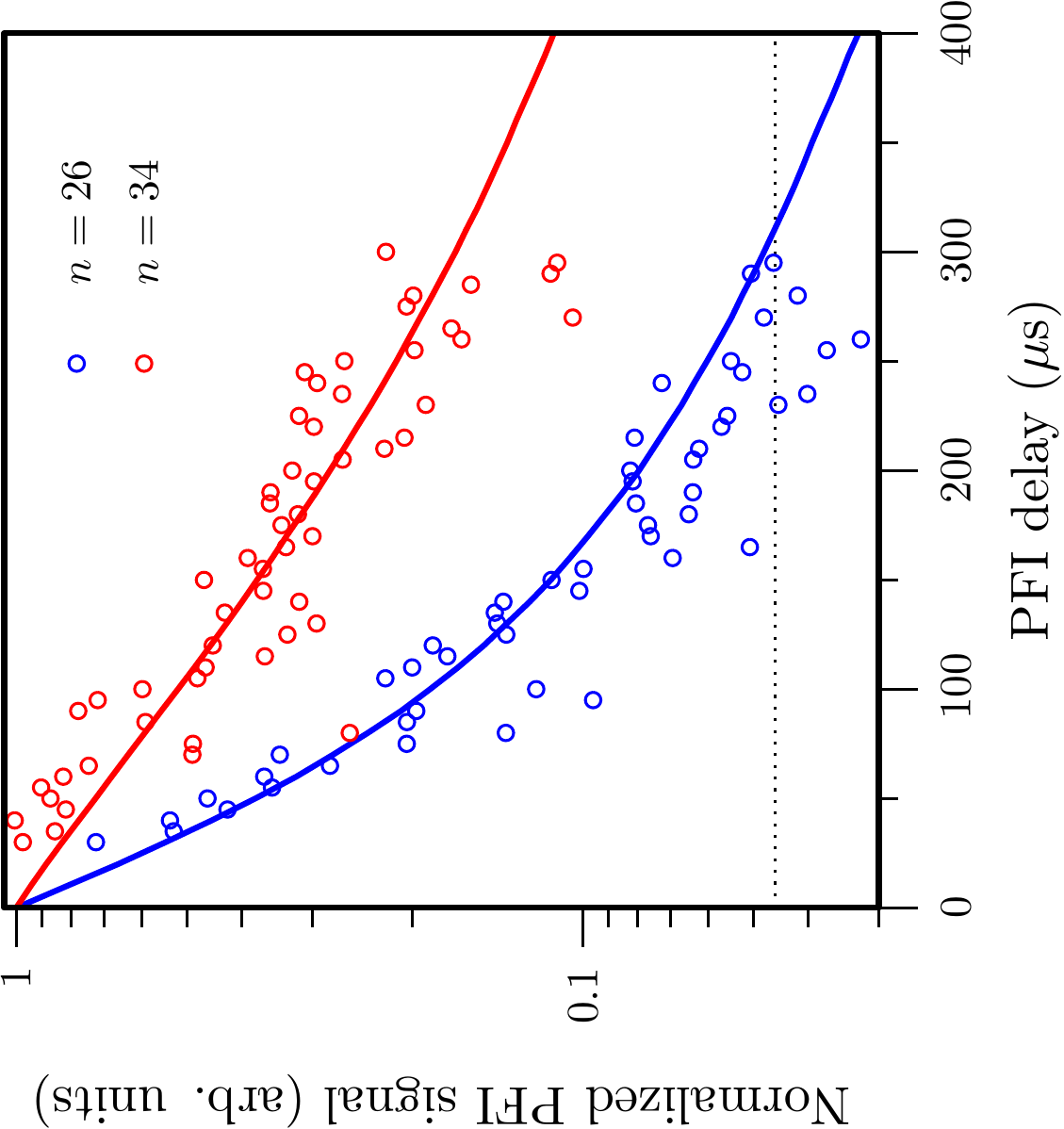}
\caption{\label{RT-decay} Comparison of experimental trap decay curves measured experimentally (open circles) at room temperature following deceleration and trapping of Rydberg H atoms of $n=26$ (blue) and $n=34$ (red) with trap-decay curves obtained from our Monte-Carlo simulations (solid lines of the corresponding color). The dotted horizontal line indicates the background signal level.}
\end{figure}

Fig.~\ref{RT-decay} compares the trap decay curves measured at room-temperature following deceleration and trapping of Rydberg H atoms of $n=26$ (blue color) and $n=34$ (red) with trap-decay curves obtained from our Monte-Carlo simulations. Although the experimental data show a larger scatter than at lower temperatures because of the larger background pressure, the calculated and measured trap-decay curves are in good agreement. To estimate the rate of black-body-radiation-induced ionization, we carried out simulations for different values and found the best agreement with the experimental data when assuming that it amounts to 10\% of the total rate of black-body-radiation-induced bound-bound transitions, which is compatible with the computed ionization rates reported in Ref.~\cite{glukhov10a}.

\begin{figure}[!tbh]
\centering
\includegraphics[angle=0,width=0.45\textwidth]{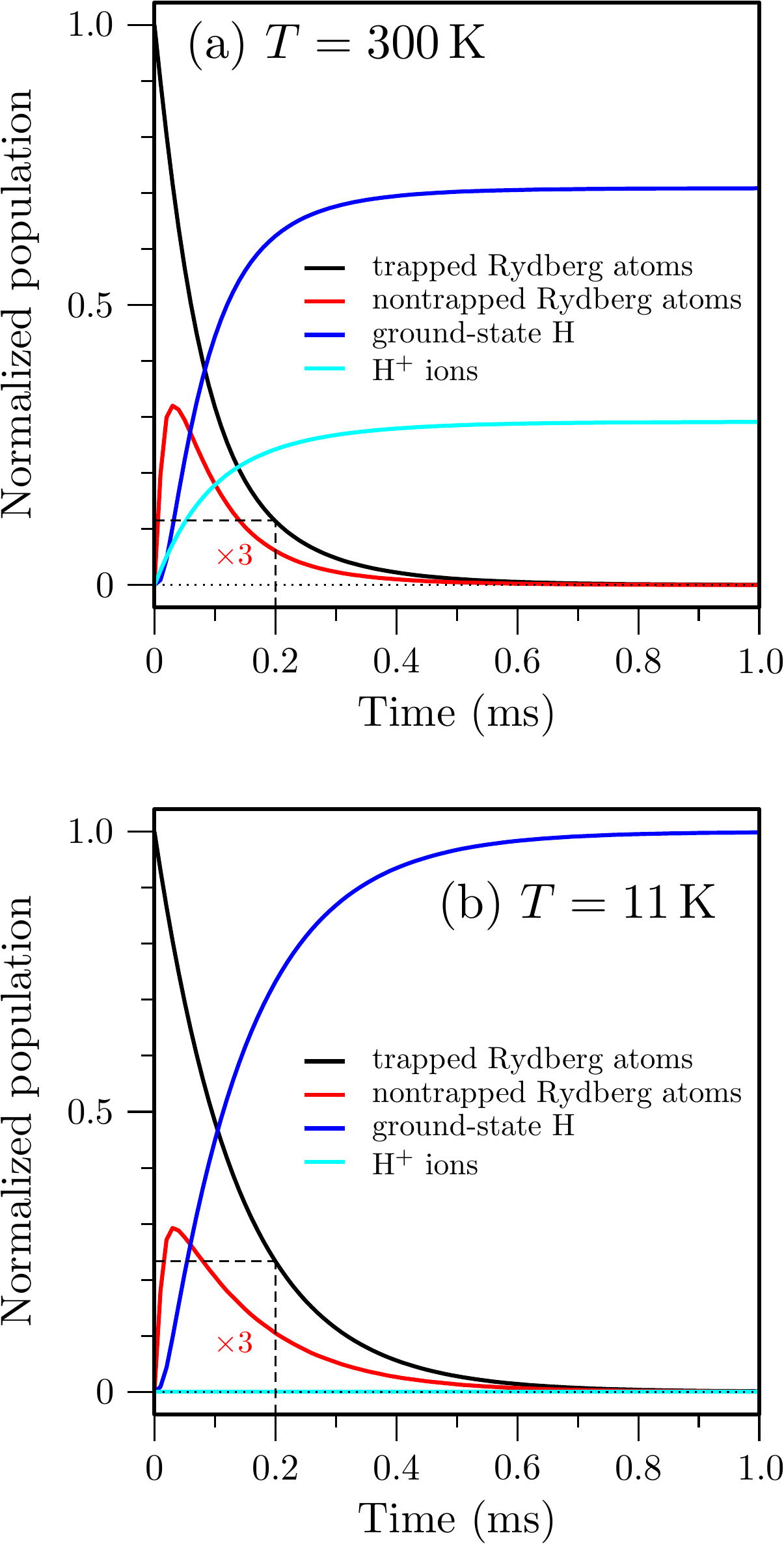}
\caption{\label{fig:population}(a,b)~Simulation of the number of Rydberg-Stark states in the electrostatic trap (black), the number of atoms in Rydberg-Stark states not fulfilling the trapping conditions (red, enhanced by a factor of 3 for better visibility), the number of 1~$^2{\rm S}{1/2}$ ground state atoms (blue), and the number of ions (cyan) as a function of time. The initial population of $n=30$ H atoms  are assumed to be distributed over $k=15-23, m=0,\pm 2$ Rydberg-Stark states, and the temperature of the radiation field is assumed to be (a)~$T=300$\,K and (b)~$T=11$\,K.}
\end{figure}

The results of a typical set of calculations of trap-decay curves are displayed in Fig.~\ref{fig:population}(a) and (b), which correspond to fictive experiments in which $n=30,k=15-23,m=0,\pm 2$ Rydberg-Stark states are initially loaded into the trap and the distribution of population is allowed to evolve for 1\,ms in thermal radiation fields corresponding to $T=300$\,K and 11\,K, respectively. In both panels, the relative population of trapped Rydberg atoms is displayed in black, the population of atoms that have decayed to the 1~$^2{\rm S}{1/2}$ ground state in blue, the ionized fraction in cyan, and the transient population in Rydberg states that do not fulfill the trapping condition in red.

The main difference between the two panels concerns the final yield of ground-state and ionized atoms, which are 71\% and 29\%, respectively at 300\,K, and 99\% and 0\% at 11\,K. The relative population of trapped Rydberg atoms after 1 ms is negligible at 300\,K but represents 1\% of the initial population at 11\,K. The main reason for this difference lies in the fact that the rates of transitions to $n < 23$ Rydberg states and
 of blackbody-radiation-induced ionization are large at  300\,K (about 1900~s$^{-1}$ for ionization) but zero at 11\,K.
 A reduction of the temperature to $T=11$\,K thus leads to an enhancement of trapped Rydberg atoms (illustrated by the dashed black lines at $200\,\mu$s in Fig.~\ref{fig:population}(a) and (b)). The dominant decay mechanism at $T=11$\,K is fluorescence, resulting in an almost complete transfer of the initially excited Rydberg atoms to the $^2{\rm S}{1/2}$ ground state within 1\,ms. The loss of atoms from the trap resulting from transitions to Rydberg-Stark states that do not have a dipole moment (red curves) sufficient for trapping is higher at 300\,K than at 11\,K, but the difference is small because fluorescence to Rydberg-Stark states with $n\leq 20$ is the dominant contribution to the trap decay.

Although the yield of ground-state atoms after 1 ms approaches 100\% at 11\,K but only 71\% at 300\,K, the initial repopulation of the ground state occurs more rapidly at  300\,K,
because stimulated emission helps to reduce $n$ and $\left|m\right|$ (see discussion at the end of subsection~\ref{subsc:blackbody-radiation}). This observation is relevant in the context of the AEGIS experiment on antihydrogen at CERN \cite{kellerbauer08a}, in which not only the yield of ground-state atoms produced from the radiative decay of Rydberg states needs to be optimized, but also the rate at which this happens is important. To maximize the yield of ground-state atom at a short time after antihydrogen production, it may be beneficial to expose the sample of antihydrogen Rydberg atoms to thermal radiation corresponding to a higher temperature than provided by the cryogenic environment of the experiment, or to intentionally illuminate the antihydrogen sample with a broadband low-frequency radiation source.

\begin{figure}[!tbh]
\centering
\includegraphics[angle=270,width=0.6\textwidth]{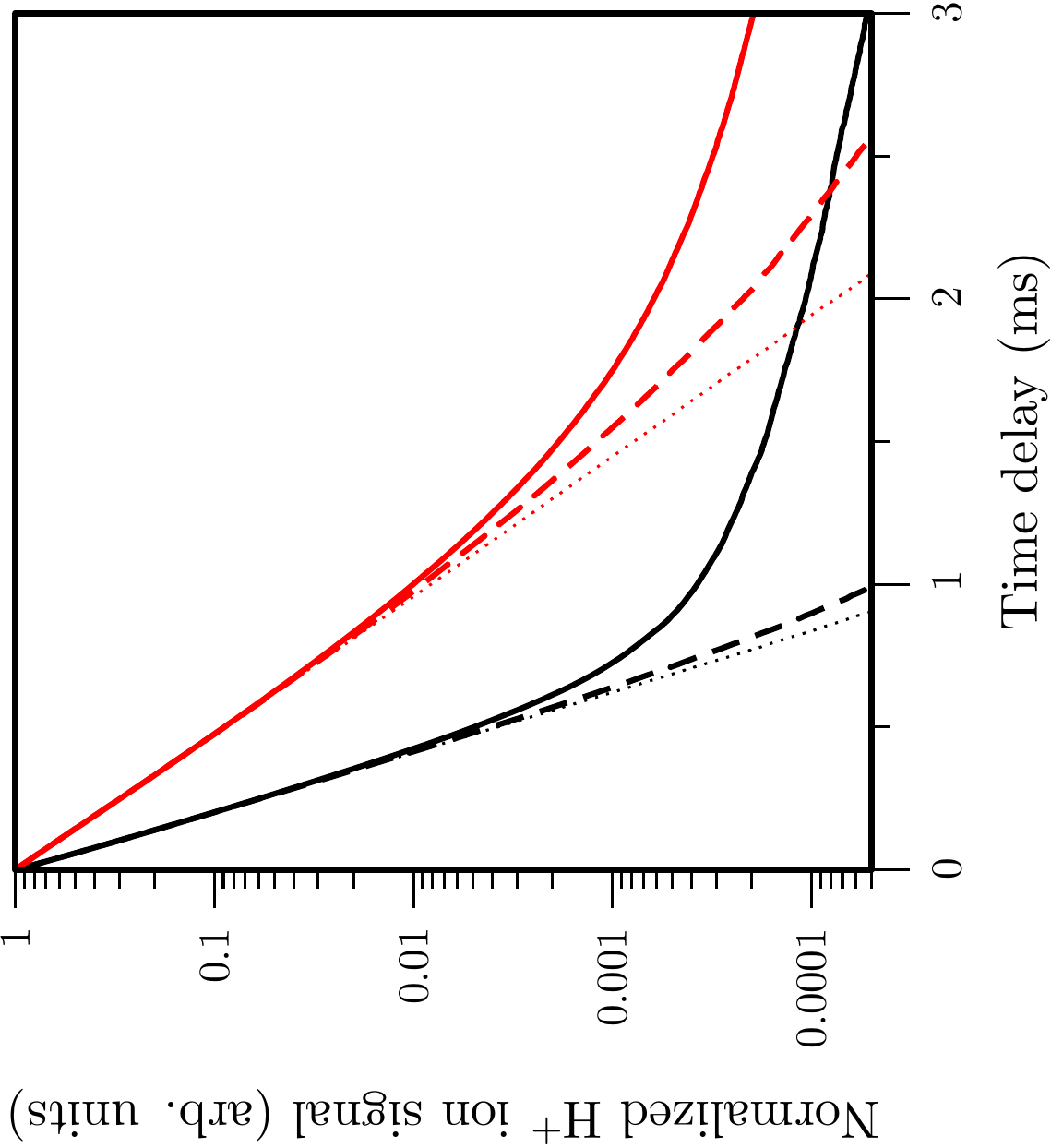}
\caption{\label{fig:ch6-T11-vs-Tmix}Monte-Carlo simulations of H atom loss from the trap after initial excitation to the $n=27$ (black) and $n=33$ (red) Rydberg states. The fluorescence decay is illustrated as dotted lines. The dashed and solid lines correspond to a blackbody temperature of 11\,K and to a blackbody-radiation field which consists of a 99\% contribution at 11\,K  and of a 1\% contribution at 300\,K, respectively.}
\end{figure}

The measurements of trap-decay curves measured after cooling the trap to 11\,K and  presented in Fig.~\ref{fig:ch6-n-dependence-experiment} had to be performed under conditions where the Rydberg sample was exposed to a small amount of room-temperature radiation (1\% solid angle) passing through the holes made for the gas and laser beams and for the ion extraction (see discussion at the end of Section~\ref{subsc:cryo-setup}). The effect of this 1\% contribution on the expected trap-decay curves can be examined by comparing Monte-Carlo simulations carried out for the $n=27$ and 33 Rydberg-Stark states used in the experiments. The relevant data are presented in Fig.~\ref{fig:ch6-T11-vs-Tmix} in black ($n=27$) and red ($n=33$) color. In both cases, the dotted, dashed, and full lines represent the calculated spontaneous decay, the decay calculated for thermal radiation at 11\,K, and for a radiation field consisting of a 99\% component at 11\,K and 1\% component at 300\,K, respectively.
A pure 11\,K radiation field only leads to a very slight deviation, visible after $800\,\mu$s (1.5\,ms) for $n=27$ ($n=33$), from the expected spontaneous decay. This deviation originates from stimulated $\left|\Delta n\right| = 1$ transitions to Rydberg-Stark states with $\left|m\right|\ge 3$, which have longer fluorescence lifetimes ($\ge 1$~ms) than the optically accessible $m=0,\pm 2$ states. Such transitions are enhanced in the presence of a small (1\%) room-temperature contribution to the radiation field and lead to a stronger deviation from the exponential decay by fluorescence. This effect is more pronounced at $n=27$ than at $n=33$ because the blackbody-radiation distribution at 300\,K peaks at a frequency corresponding to a $\Delta n=\pm 1$ transition at $n\sim 20$. Overall, the effect remains small: After a time of 2\,ms, a fraction of only $1/10000$ ($1/1000$) of the initially prepared $n=27$ ($n=33$) states has been redistributed to Rydberg states which have lifetimes longer than $1$\,ms.

Figures~\ref{fig:ch6-n-dependence-experiment}(a,b) compare the results of Monte-Carlo simulations with the results of trap-decay measurements carried out with H Rydberg-Stark states in the range of principal quantum number between 23 and 33. In this figure, the dashed lines correspond to the fluorescence lifetimes of the respective states calculated with Eq.~(\ref{eq:tau-parabolic}) which are $86\,\mu$s, $103\,\mu$s, $118\,\mu$s, $138\,\mu$s, $157\,\mu$s, $181\,\mu$s and $207\,\mu$s for the $n=27$, 28, 29, 30, 31, 32 and 33 Stark states, respectively. The solid lines correspond to Monte-Carlo simulations including all radiative decay processes, i.e., fluorescence (Eq.~(\ref{eq:tau-parabolic})), blackbody-radiation-induced ionization and blackbody-radiation-induced $n$-changing transitions (Eqs.~(\ref{eq:Einstein-B}) and (\ref{eq:total-decay}). The simulations were carried out for a radiation field at 11\,K with a 1\% contribution of room-temperature radiation, corresponding to the experimental conditions. At early times up to about 500\,$\mu$s, the calculations are in good agreement with the simulations. In this region, the decay is primarily caused by fluorescence, and the rare stimulated transitions to long-lived Rydberg states are not visible yet. Their effects become visible as a deviation from the initial exponential behavior when the population of trapped atoms has decayed to less than 10\% of the initial value.

At this point, the calculated trap-decay curves also begin to markedly deviate from the experimentally observed decay curves and indicate that more long-lived Rydberg states are formed in the experiment than the Monte-Carlo simulations predict.  After 2\,ms, the calculated amount of trapped atoms is almost two orders of magnitude less than observed experimentally. The slopes of the calculated and experimental decay curves after 2 ms, however, are in good agreement, and correspond to a lifetime of $\approx 1.2$\,ms, which corresponds to the radiative lifetime of an $n=51$, $\left|m\right|=2$ Rydberg-Stark state or of an ensemble of states with lower $n$ but higher $\left|m\right|$ values. The rates that would be necessary to induce multiple $\Delta n$-transitions to states of $n\approx 50$ are much larger than predicted by Eq.~(\ref{eq:Einstein-B}) and this mechanism can be ruled out. The two-order-of-magnitude enhancement of the yield of long-lived Rydberg states observed experimentally must therefore be the result of a more effective mechanism to populate levels with $\left|m\right|>2$ than included in our model. Such an enhancement could for instance result from a much larger contribution (more than 15\%) of room-temperature radiation than the 1\% assumed in the calculations. A much lager room-temperature contribution would, however, increase the rate of blackbody-radiation-induced ionization and cause a faster decay at early times.  The good agreement between the simulations and experimental results at early times (see Fig.~\ref{fig:ch6-n-dependence-experiment}(a,b)) and the measures taken to verify that room-temperature radiation does not significantly affect the observed trap-decay curves (see Section~\ref{subsc:cryo-setup})
enable us to also rule out this explanation. Collisions between the trapped Rydberg atoms might contribute to induce transitions to levels with $\left|m\right|>2$  and the modeling of such collisions is presented in the next section.

\section{\label{sec:ch6-theory} Collision-induced $m$-changing processes: Theoretical considerations and simulations}

\subsection{\label{resonant_collisions} Collision-induced transitions between near-degenerate levels of pairs of atoms}
\noindent
In this section we examine the potential role of collisions between Rydberg atoms as a possible mechanism to account for the deviations between measured trap-decay curves obtained following off-axis trapping of hydrogen atoms in Rydberg-Stark states and the Monte-Carlo simulations of these curves based on a treatment of radiative processes.
The dominant term in the long-range electrostatic interaction series between two Rydberg atoms is the dipole-dipole interaction term \cite{flannery05a}. This interaction is at the origin of diverse phenomena in cold Rydberg atom gases, such as resonant transitions of Rydberg-atom pairs \cite{safinya81a,pillet87a}, inelastic collisions \cite{gallagher94a}, Penning ionization \cite{viteau08a,reinhard08a}, Rydberg-excitation blockade \cite{singer04a,tong04a,vogt06a} and the formation of Rydberg macrodimers \cite{farooqi03a,overstreet07a,sassmannshausen15a}. It also plays an important role in the evolution of a cold Rydberg gas into a plasma \cite{vitrant82a,robinson00a,pohl03a}.

The Rydberg atom densities in our experiments (in the range 10$^5$-10$^6$ Rydberg atoms/cm$^{3}$) are lower than in most experiments devoted to studies of dipole-dipole interactions and which can reach densities of more than 10$^{10}$ Rydberg atoms/cm$^{3}$ \cite{raimond81a,gallagher94a,farooqi03a,vogt07a,overstreet09a,samboy11a,sassmannshausen15a}. The presence of the trapping electric fields in our experiments implies that the collisions are between atoms in $\left|nkm\right>$ Stark states rather than between atoms in $\left|n\ell m\right>$ states. Because atoms only interact if they are in close proximity, one can assume, to a good approximation, that the interacting atoms are subject to electric fields of similar strengths.

\begin{figure*}[!h]
\centering
\includegraphics[angle=0,width=1.0\textwidth]{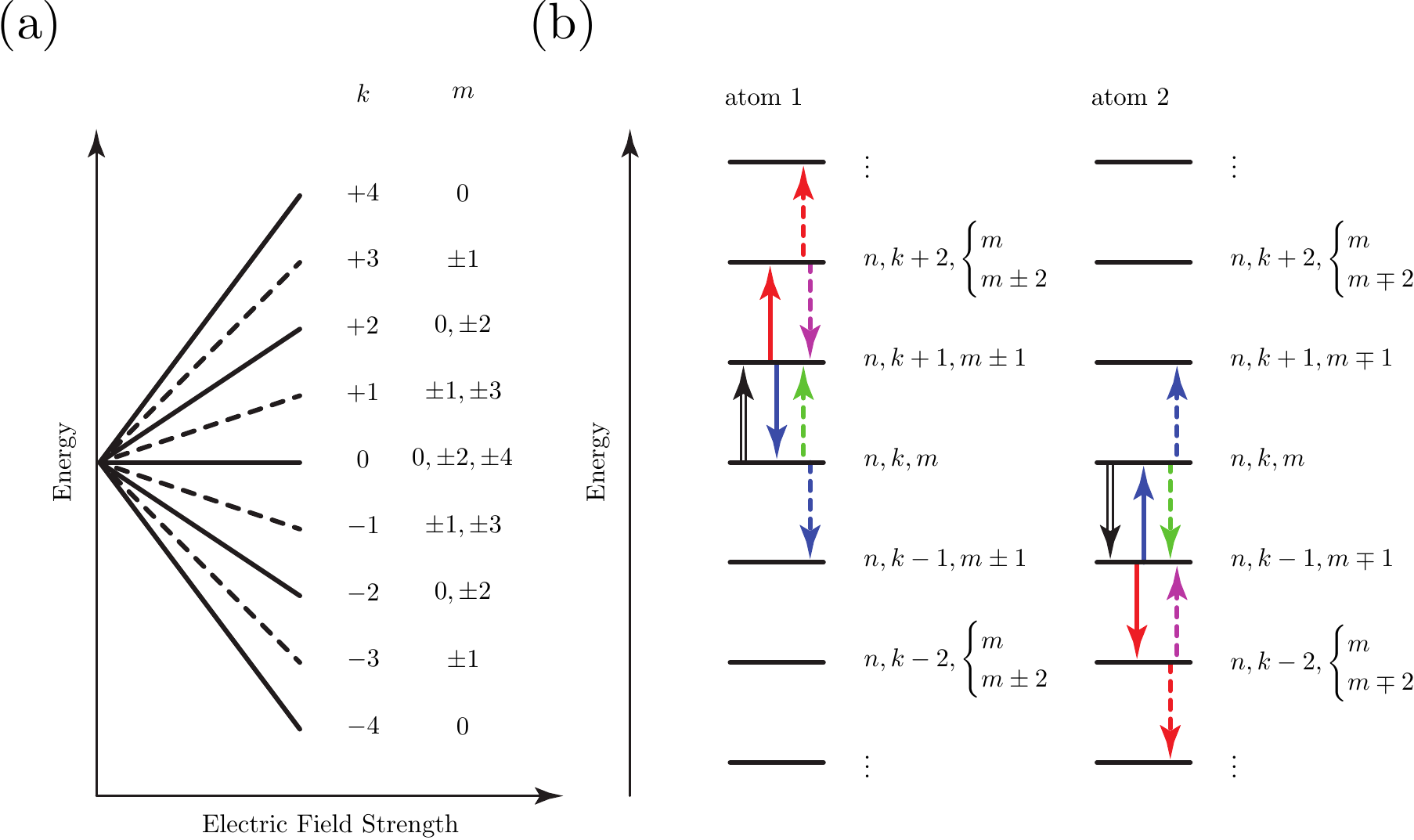}
\caption{(a)~Energy level structure of the $n=5$ Rydberg-Stark states of atomic hydrogen (adapted from Ref.~\cite{merkt94b}). (b)~Schematic diagram depicting parts of the network of states connected by the resonant dipole-dipole interaction. Pairs of arrows of the same shape or colour indicate the transitions of the two-atom system.}
\label{fig:stark-levels}
\end{figure*}
In the case of Rydberg states of the H atom with $n$ values around 30 and for the relevant electric field strengths (10-50 V/cm), the Stark effect is essentially linear so that the Stark states of a given $n$ value form a manifold of energetically equidistant levels, depicted schematically in Fig.~\ref{fig:stark-levels}(a) for $n=5$. This equidistance between Stark states intrinsically leads to resonances when Rydberg atoms prepared in Stark states of the same $n$ manifold collide, because the energy needed to excite one of the two atoms (index 1) to the next higher Stark state ($\Delta k_1= +1$) is almost exactly compensated if the other atom (index 2) undergoes a transition to the next lower Stark state ($\Delta k_2= -1$). The dominant nature of the dipole-dipole interaction term in the long-range interaction series and the resonant nature of the collisions make transitions of the type
\begin{equation}\label{dipdip}
\left|nk_1m_1\right>\left|nk_2m_2\right> \rightarrow \left|nk_1\pm 1m_1'\right>\left|nk_2\mp 1m_2'\right>,
\end{equation}
by far the most important  of all collision-induced processes in our electrostatic traps. We therefore limit our treatment of collision-induced transitions to such transitions.

To include these transitions into our Monte-Carlo simulations of the trap-decay curves, we use a hard-sphere collision model in combination with second-order kinetics. The rate of a transition $\left|nk_{1,i}m_{1,i}\right>\left|nk_{2,i}m_{2,i}\right> \rightarrow \left|nk_{1,j}m_{1,j}\right>\left|nk_{2,j}m_{2,j}\right>$ of the kind described by Eq.~(\ref{dipdip}) is given by
\begin{equation}\label{rate-eq}
-\frac{\text{d} N_{1,i}(t)}{\text{d}t} =  -\frac{\text{d} N_{2,i}(t)}{\text{d}t}  = k_{ij}N_{1,i}(t) N_{2,i}(t),
\end{equation}
where $N_{1,i}(t)$ and $N_{2,i}(t)$ are the number densities of the colliding atoms (index 1 and 2) in their respective Stark states prior to the collisions (index $i$). The rate constant $k_{ij}$ for a hard-sphere collision is
\begin{equation}
k_{ij} = \sqrt{\frac{8k_{\text{B}}T}{\pi\mu_{\rm R}}}\,\sigma_{ij} = \bar{v}_{\rm rel} \,\sigma_{ij} .
\label{eq:2}
\end{equation}
In the case considered here, the reduced mass $\mu_{\rm R}$ is $m_{\rm H}/2$ and
\begin{equation}
\sigma_{ij} = \frac{1}{4\pi\varepsilon_0}\frac{\pi^2}{2\hbar}\,\frac{\left<n,k_{1,i},m_{1,i}\left|\hat{\mu}\right|n,k_{1,j},m_{1, j}\right> \left<n,k_{2,i},m_{2,i}\left|\hat{\mu}\right|n,k_{2,j},m_{2,j}\right>}{\bar{v}_{\rm rel}}=\pi r^2_{\rm hs}~,
\label{eq:3}
\end{equation}
is the cross section for the resonant collision and can be calculated using the expressions for the electric-dipole transitions presented in Subsection~\ref{subsc:spontaneous-emission}.
In Eqs.~(\ref{eq:2}) and ~(\ref{eq:3}),
$ \bar{v}_{\rm rel}$
is the average relative velocity of a thermal gas, i.e., $\approx 80$~m/s at the 150~mK translational temperature of the trapped atoms and $r_{\rm hs}$ is an effective hard-sphere radius.

Starting with a sample consisting of atoms being all in a $\left| n,k,m \right>$ Rydberg-Stark state prepared optically, only four resonant transitions (labeled (i)-(iv)) are of the type specified by Eq.~(\ref{dipdip})
\begingroup\footnotesize
\begin{eqnarray*}
\left|n,k,m\right>\left|n,k,m\right> &\stackrel{{\rm (i)}}{\longrightarrow}& \left|n,k+1,m+1\right>\left|n,k-1,m-1\right> \\
\left|n,k,m\right>\left|n,k,m\right> &\stackrel{{\rm (ii)}}{\longrightarrow}& \left|n,k+1,m-1\right>\left|n,k-1,m+1\right> \\
\left|n,k,m\right>\left|n,k,m\right> &\stackrel{{\rm (iii)}}{\longrightarrow}& \left|n,k-1,m+1\right>\left|n,k+1,m-1\right> \\
\left|n,k,m\right>\left|n,k,m\right> &\stackrel{{\rm (iv)}}{\longrightarrow}& \left|n,k-1,m-1\right>\left|n,k+1,m+1\right>.
\end{eqnarray*}
\endgroup
and have large cross sections. This dominance results from the conservation of the quantum number $M=m_1+m_2$ and the stronger nature of the $\Delta k = \pm 1$
dipole-dipole transitions compared to $\Delta k = \pm 2,\pm 3,\ldots$ transitions. For a Rydberg-atom sample initially excited to the $\left|32,27,0\right>$ state, these four channels are:
\begin{eqnarray*}
\left|32,27,0\right>\left|32,27,0\right> &\stackrel{{\rm (i)}}{\longrightarrow}& \left|32,26,+1\right>\left|32,28,-1\right> \\
\left|32,27,0\right>\left|32,27,0\right> &\stackrel{{\rm (ii)}}{\longrightarrow}& \left|32,26,-1\right>\left|32,28,+1\right> \\
\left|32,27,0\right>\left|32,27,0\right> &\stackrel{{\rm (iii)}}{\longrightarrow}& \left|32,28,+1\right>\left|32,26,-1\right> \\
\left|32,27,0\right>\left|32,27,0\right> &\stackrel{{\rm (iv)}}{\longrightarrow}& \left|32,28,-1\right>\left|32,26,+1\right>.
\end{eqnarray*}

We consider a collision to be resonant if the collisional broadening
\begin{equation} \label{gamma_coll}
\Gamma_{\text{coll}}  = \frac{\Delta E}{hc} \approx \frac{\mu_1\mu_2}{4\pi\varepsilon_0\,hc r^3_{\rm hs}}
\end{equation}
is larger than the energy change resulting from the transition. If the interacting atoms are subject to exactly the same field strength (in the range between 10~V/cm and 46~V/cm for our trap), the energy change only arises from the slight differences in the quadratic Stark shifts of the initial and final states, i.e., $\Delta E_{ij,\rm Stark}^{(2)}(F) = 1.5 n^4F^2(a_0e)^2/E_{\rm h}$. In the case of $n=30,m=0$ Stark states, the hard-sphere radius $r_{\rm hs}$ is about 10~$\mu$m and $\Gamma_{\text{coll}}$ is $2.7\cdot 10^{-5}$~cm$^{-1}$, which is of the same order of magnitude as $\Delta E_{ij,\rm Stark}^{(2)}$ in the relevant range of electric-field strengths. In general, however, interacting atoms separated by a distance $r$ experience slightly different electric-field strengths because of the inhomogeneity of the trapping field so that there is a first-order contribution
\begin{equation} \label{Stark_detuning}
\Delta E_{ij,\rm Stark}^{(1)}(F) = 1.5 \left(\frac{\partial F}{\partial r}\right)    a_0en  r
\end{equation}
to the energy change. Taking into account the typical field gradient $(\partial F/\partial r)$ of $1.2\cdot 10^{6}$~V/m$^2$ in the trap, one obtains $\Delta E_{ij,\rm Stark}^{(1)}(F) /(hc)= 2.3 \cdot 10^{-4}$~cm$^{-1}$  at a distance of 10~$\mu$m corresponding to the hard-sphere radius of an $n=30$ Stark state, which is much larger than the estimated collisional broadening and effectively breaks the resonance condition and reduces the values of the hard-sphere radii.  The reduced hard-sphere radii and cross sections can be derived by equating the expressions for $\Gamma_{\text{coll}}$ (Eq.~(\ref{gamma_coll})) and $\Delta E_{ij,\rm Stark}^{(1)}(F)$ (Eq.~(\ref{Stark_detuning})), which, with the field gradient given above, leads to
\begin{equation}\label{hs_new}
 r_{\rm hs} = 4.6\cdot 10^{-7}n^{3/4}~{\rm m}.
\end{equation}
The reduced hard-sphere radii are about 5.4, 5.9, and 6.3~$\mu$m at $n=27$, 30, and 33, respectively, and the corresponding hard-sphere cross sections are $9.2 \cdot 10^{-11}$, $1.1 \cdot 10^{-10}$, and $1.25 \cdot 10^{-10}$~m$^2$, i.e., a factor of about two to four less than estimated with Eq.~(\ref{eq:3}).

Because of the equal spacing between Rydberg-Stark states of a given $n$-value (see Fig.~\ref{fig:stark-levels}(a)), the resonance condition also applies to further transitions obeying the selection rules $\Delta m_1=-\Delta m_2=\pm 1$, and $\Delta k_1=-\Delta k_2=\pm i (i=1,2,\ldots )$, as described schematically in Fig.~\ref{fig:stark-levels}(b).
The network of states connected by successive resonant dipole-dipole transitions is represented by Eq.~(\ref{eq:15}) in the notation~$\binom{k_1,m_1}{k_2,m_2}$, where the top and bottom lines refer to the first and second atoms, respectively
\begin{equation}
\begin{bmatrix}
\binom{k,+m_{\text{max}}}{k,-m_{\text{max}}} & \leftrightarrows & \cdots & & \cdots & & \cdots & \leftrightarrows & \binom{k+m_{\text{max}},0}{k-m_{\text{max}},0} \\
\updownarrows & & \updownarrows & & \updownarrows & & \updownarrows & & \updownarrows \\
\cdots & \leftrightarrows & \binom{k,+2}{k,-2} & \leftrightarrows & \binom{k+1,+1}{k-1,-1} & \leftrightarrows & \binom{k+2,0}{k-2,0} & \leftrightarrows & \cdots \\
 & & \updownarrows & & \updownarrows & & \updownarrows & & \\
\cdots & \leftrightarrows & \binom{k-1,+1}{k+1,-1} & \leftrightarrows & \binom{k,0}{k,0} & \leftrightarrows & \binom{k+1,-1}{k-1,+1} & \leftrightarrows & \cdots \\
 & & \updownarrows & & \updownarrows & & \updownarrows & & \\
\cdots & \leftrightarrows & \binom{k-2,0}{k+2,0} & \leftrightarrows & \binom{k-1,-1}{k+1,+1} & \leftrightarrows & \binom{k,-2}{k,+2} & \leftrightarrows & \cdots \\
\updownarrows & & \updownarrows & & \updownarrows & & \updownarrows & & \updownarrows \\
\binom{k-m_{\text{max}},0}{k+m_{\text{max}},0} & \leftrightarrows & \cdots & & \cdots & & \cdots & \leftrightarrows & \binom{k,-m_{\text{max}}}{k,+m_{\text{max}}} \\
\end{bmatrix}.
\label{eq:15}
\end{equation}
One can deduce from Eq.~(\ref{eq:15}) that $m_{\text{max}}=n-|k|-1$. For the initial state $n=32$, $k=27$, $m=0$, used as example above, Eq.~(\ref{eq:15}) becomes
\begin{equation}
\begin{bmatrix}
\binom{27,+4}{27,-4}	& \binom{28,+3}{26,-3}	& \binom{29,+2}{25,-2}	& \binom{30,+1}{24,-1}	& \binom{31,0}{23,0}	\\
\binom{26,+3}{28,-3}	& \binom{27,+2}{27,-2}	& \binom{28,+1}{26,-1}	& \binom{29,0}{25,0}	& \binom{30,-1}{24,+1}	\\
\binom{25,+2}{29,-2}	& \binom{26,+1}{28,-1}	& \binom{27,0}{27,0}	& \binom{28,-1}{26,+1}	& \binom{29,-2}{25,+2}	\\
\binom{24,+1}{30,-1}	& \binom{25,0}{29,0}	& \binom{26,-1}{28,+1}	& \binom{27,-2}{27,+2}	& \binom{28,-3}{26,+3}	\\
\binom{23,0}{31,0}	& \binom{24,-1}{30,+1}	& \binom{25,-2}{29,+2}	& \binom{26,-3}{28,+3}	& \binom{27,-4}{27,+4}	\\
\end{bmatrix}.
\label{eq:16}
\end{equation}
Similar networks can be  derived in analogy for collisions initially in Stark states of different $m$ and $k$ values.

To model the evolution of the population within the network of near-degenerate states induced by dipole-dipole collisions, we assume that the resonant collision induces transitions to all $(m_{\rm max}+1)^2$ near-degenerate states with equal probabilities and thus equal rates, which simplifies the inclusion of collision-induced processes in our Monte-Carlo simulations of trap-decay curves. Another simplification resulting from these considerations is that the cross sections and the rates for state-changing collisions only depend on $n$ and not on $k$.

\subsection{\label{subsc:ch6-monte-carlo}Monte-Carlo simulations including resonant dipole-dipole collisions between trapped Rydberg-Stark states}

The low initial density ($\approx 10^6$ cm$^{-3}$) of the trapped atoms and its rapid initial decay by radiative processes imply the approximate validity of a single-collision treatment, which, in turn, enables one to only consider collisions involving atoms in the initially prepared states.
Eq.~(\ref{eq:ch6-master-equation})
can then be used to describe the initial decay of the levels prepared optically
\begin{equation}
-\frac{\text{d} N_i}{\text{d}t} = K_i^{\text{tot}} N_i + \sum_j k_{\text{tot},ij} N_iN_j.
\label{eq:ch6-master-equation}
\end{equation}
In Eq.~(\ref{eq:ch6-master-equation}), $K_i^{\text{tot}}$ (see Eq.~(\ref{eq:total-decay})) accounts for the radiative transitions and $k_{\text{tot},ij}$, evaluated as $\bar{v}_{\rm rel}\sigma_{ij, {\rm hs}}=\bar{v}_{\rm rel}\pi r_{\rm hs}^2$ using Eq.~(\ref{hs_new}),
represents the total bimolecular rate constant for state-changing collisions between two Rydberg atoms in states $i=\left|n_ik_{i}m_{i}\right>$ and $j=\left|n_ik_{j}m_{j}\right>$, with number densities $N_i$ and $N_j$, respectively. In our model, $k_{\text{tot},ij}$ only depends on $n$ and is equal, under our experimental conditions, to $7.4 \cdot 10^{-9}$, $8.6 \cdot 10^{-9}$, and $10^{-8}$~m$^3$/s  for $n=27$, 30 and 33, respectively. All Rydberg states with $n_j\neq n_i$ can be disregarded because the resonance condition for dipole-dipole transitions involving the atoms prepared optically is only fulfilled if both atoms are in a Stark state of the same $n$ manifold.

The assumption of equal transition probabilities to the $g$-fold
near-degenerate pair states following a resonant collision ($g=(m_{\text{max}}+1)^2$ in the case of the network in Eq.~(\ref{eq:15}))
implies that the rate constant for the formation of each of these states from the initial pair state $\left|nk_{i}m_{i}\right>\left|nk_{j}m_{j}\right>$ is $k_{\text{tot},ij}/g$.

The implementation of this simple collision model in the Monte-Carlo simulation of the trap-decay curves necessitates the extension of the flow diagram presented in Fig.~\ref{fig:flow-diagram} by the nodes marked in red.
The number density $N_0=\sum_{k,m} N_{0,k,m}$ ($N_{0,k,m}$ represents the number density of atoms in the different optically accessible $k,m$ Rydberg-Stark states specified in Subsection~\ref{subsc:laser-setup}) of initially excited atoms
is needed to calculate the bimolecular decay rates at each time interval. To this end, an initial particle number density $N_{0}$ is used as an adjustable parameter, which is optimized to obtain a good agreement between the simulation and the experimental data.
In analogy to the treatment of radiative processes described in Subsection~\ref{subsc:monte-carlo}, the probability of a collision-induced transition involving one of the trapped Rydberg atoms
\begin{equation}
P_{n_ik_im_i}= \frac{\left|\Delta N_i\right|}{N_i} = k_{\text{tot},i} N_j\Delta t
\label{eq:PnkmDeltam}
\end{equation}
is calculated at each time step. In Eq.~(\ref{eq:PnkmDeltam}), $\Delta N_i$ is the change of particle number density. To decide whether a state-changing collision has taken place or not, $P_{n_ik_im_i}$ is compared against a random probability $P_4$ ($0<P_4\leq 1$). A second random probability determines which of the $g$ near-degenerate final states is populated by the collision.
Having to keep track, at each time step, of the state of pairs of particles forces one to perform, for each particle $N_1$, a second loop over the remaining particles $N_2$ and to update the corresponding number densities, which considerably slows down the simulations compared to the treatment of pure radiative processes.

\begin{figure*}[!h]
\centering
\includegraphics[angle=270,width=1.0\textwidth]{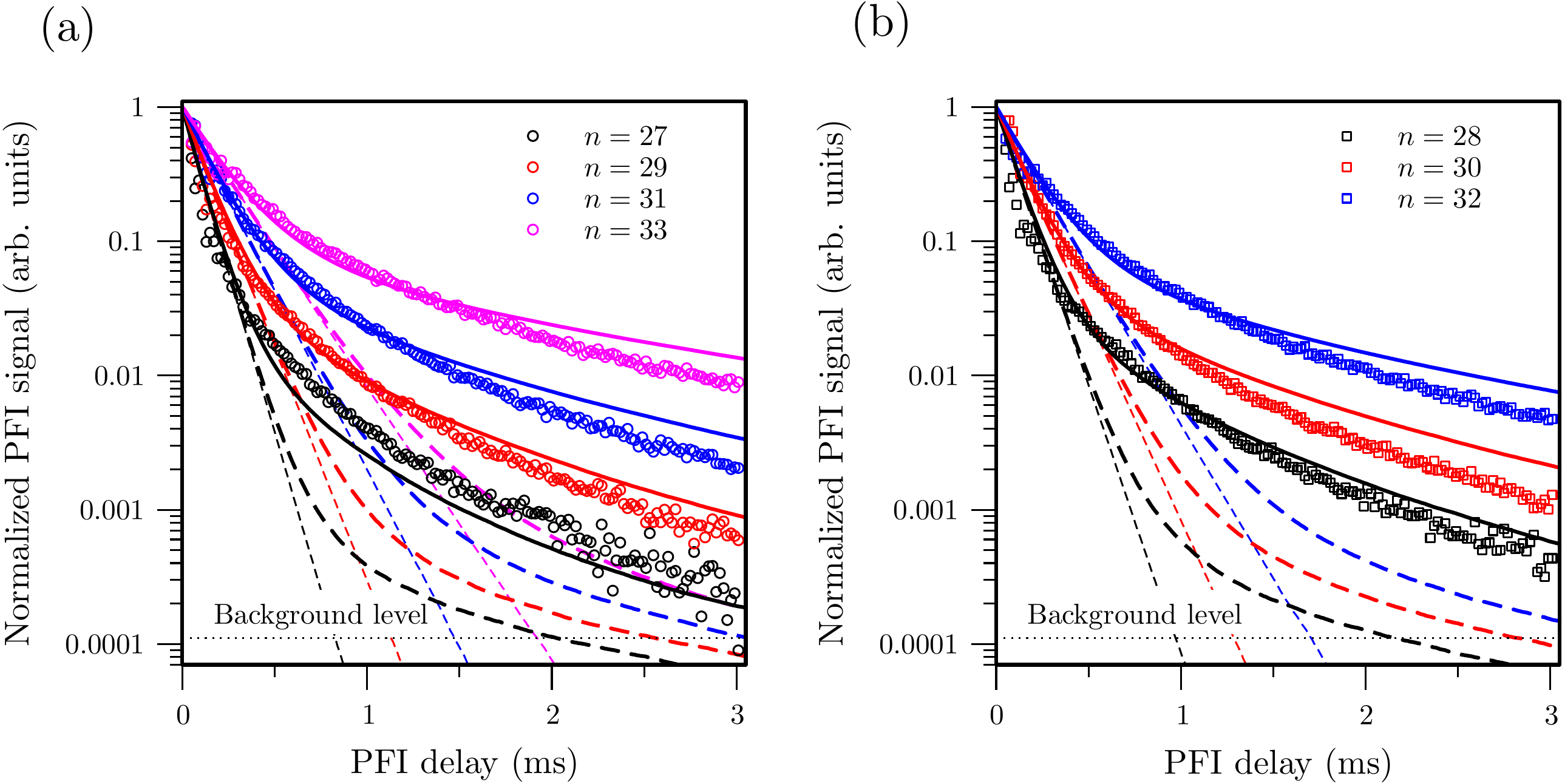}
\caption{\label{fig:ch6-simulation-comparison} Trap-decay curves recorded following deceleration and trapping of H atoms in Rydberg-Stark states of principal quantum numbers $n$ between 27 and 33 in an environment held at 11~K. The thin dashed lines represent the decay of the initially prepared states by fluorescence. The thick dashed lines correspond to the decay calculated by considering all radiative processes. The simulations displayed as solid lines include, in addition, resonant dipole-dipole collisions (see text for details).
}
\end{figure*}

\begin{figure*}[!h]
\centering
\includegraphics[angle=0,width=0.65\textwidth]{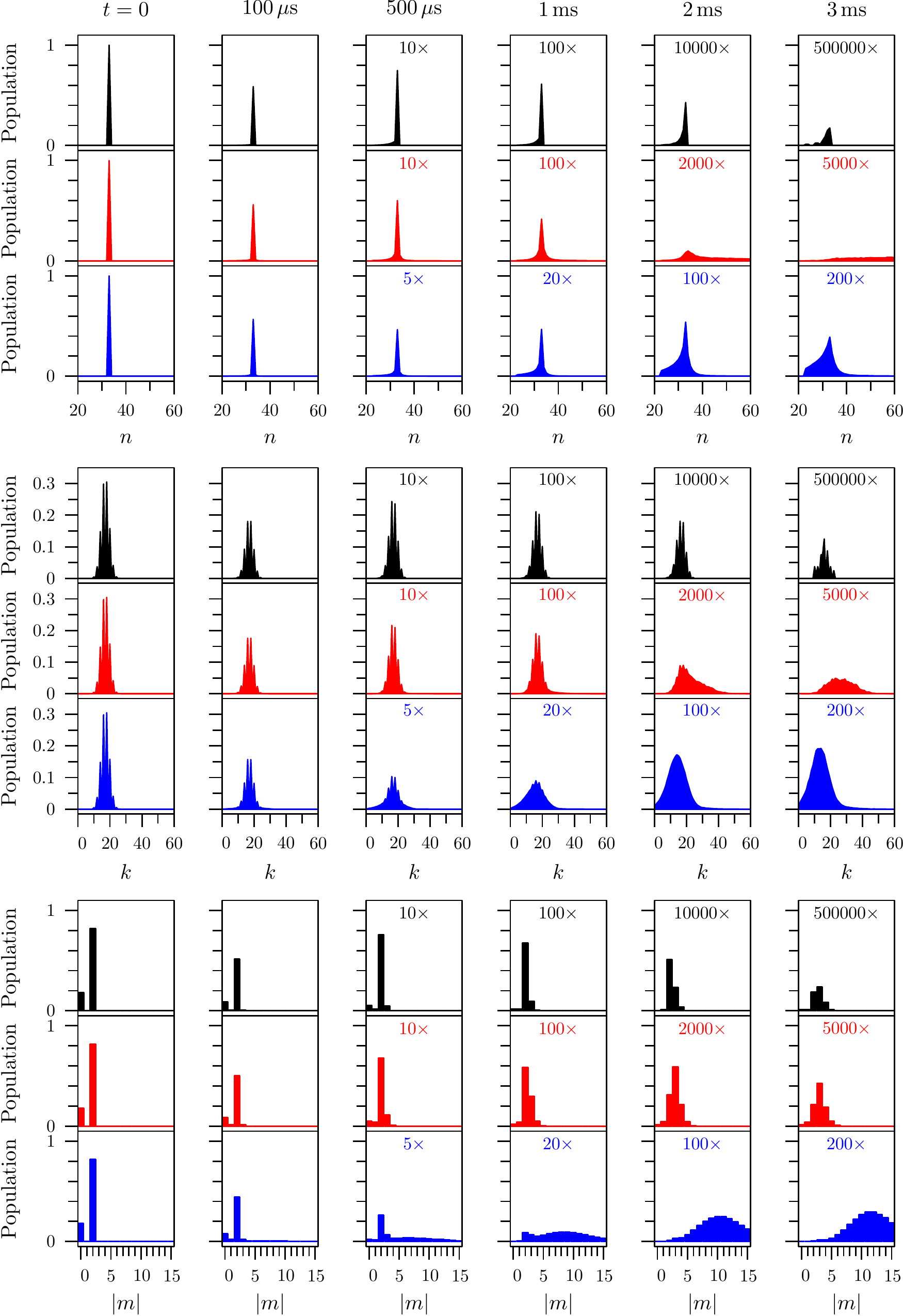}
\caption{\label{fig:ch6-population-collision} Calculated distributions of the population over states of different principal quantum number~$n$ (top row), different Stark-state index~$k$ (middle row), and different magnetic quantum number~$\left|m\right|$ (bottom row) following deceleration and trapping of H atoms in $n=33$ Rydberg-Stark states. From left to right, the different panels present the initial state distribution, and the state distributions after $t=100\,\mu$s, 500\,$\mu$s, 1\,ms, 2\,ms and 3\,ms. The thermal radiation field is assumed to consist of 99\% of the 11\,K thermal radiation field and of 1\% of the 300\,K radiation field. The distributions displayed were obtained by considering only spontaneous emission (black), all radiative processes (red), and all radiative processes and collisions (blue). The scales of the distributions calculated for the later times have been enlarged by the factors indicated in the upper right corners of the different panels.}
\end{figure*}

Figure~\ref{fig:ch6-simulation-comparison} compares the trap-decay curves measured at 11~K after trapping atoms with $n$ values between 27 ind 33 (same data as in Fig.~\ref{fig:ch6-n-dependence-experiment}) with three sets of computed decay curves. The thin dashed lines correspond to the spontaneous decay, the thick dashed lines to the Monte-Carlo simulations of all radiative processes as discussed in Section ~\ref{subsc:blackbody-radiation-tot}, and the full lines to the Monte-Carlo simulations including resonant collisions. This third set of curves was obtained for an initial Rydberg-atom number density of $2\cdot 10^{5}$\,cm$^{-3}$, which was found to yield the best agreement with the experimental results.

\begin{figure}[!tbh]
\centering
\includegraphics[angle=0,width=0.45\textwidth]{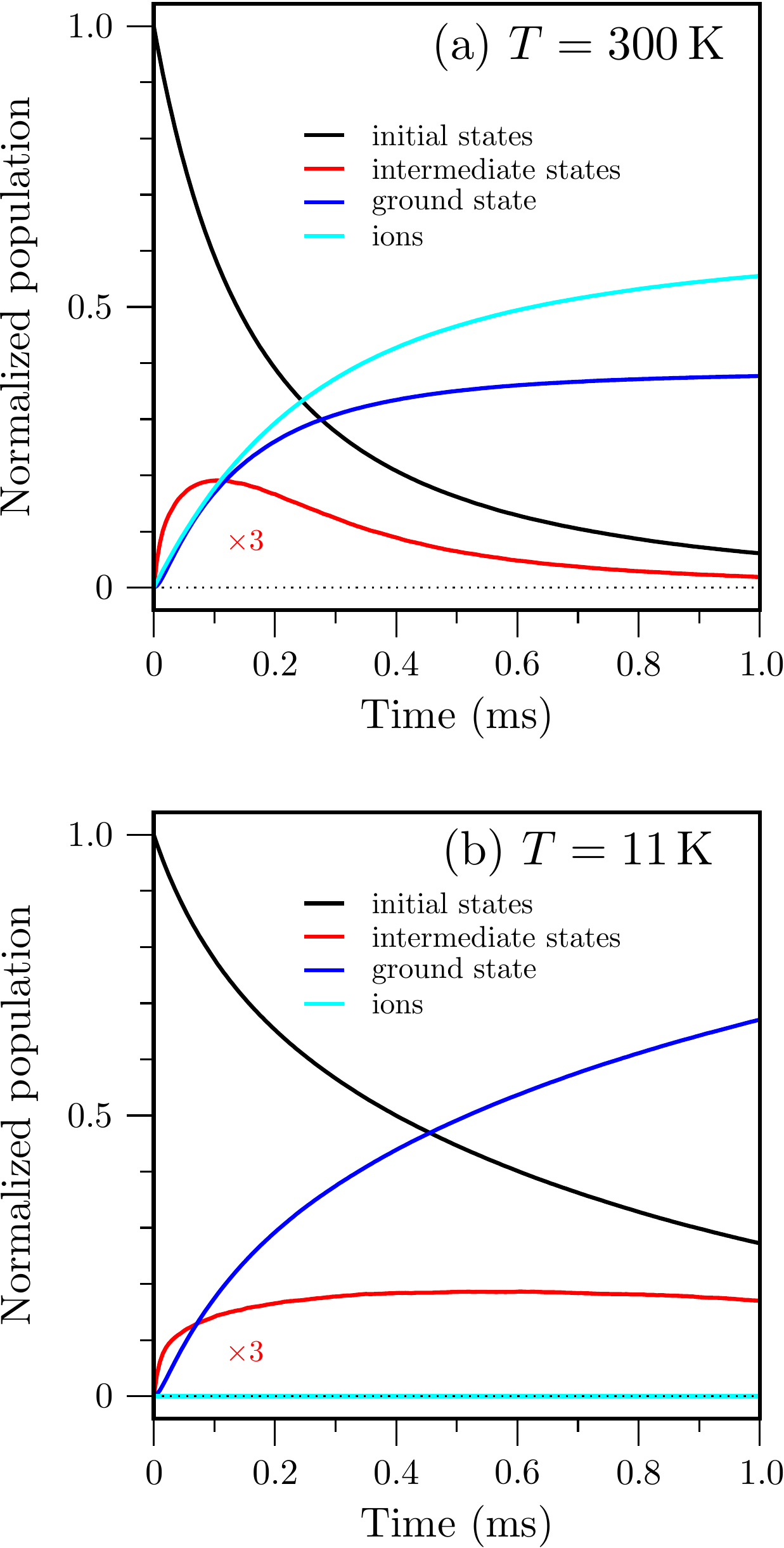}
\caption{\label{aegis}(a)~Simulations of trap-decay curves for a sample Rydberg hydrogen atoms initially equally distributed over the states $n=30$, $\left|m\right|=0$-15 obtained for blackbody-radiation temperatures of 11\,K (lower panel) and 300\,K (upper panel). The black and dark-blue lines represent the relative population in the initial states, and the ground state, respectively, and the pale-blue line corresponds to the ionized fraction. The red lines, the intensity of which has been magnified by a factor of three, corresponds to the the population of atoms in intermediate states.}
\end{figure}

The early decay is dominated by fluorescence so that the four curves presented for each value of $n$ are almost identical up to $\approx 500$\,$\mu$s. At later times, however, the trap decay deviates from the initial exponential behavior because of $\Delta n$ and $\Delta m$ transitions to longer-lived states. The curves obtained by Monte-Carlo simulations predict the right slope of the trap decay curve after 2\,ms , but the curve calculated by including only radiative processes underestimates the numbers of trapped atoms by almost two orders of magnitude, as already discussed in Section~\ref{subsc:blackbody-radiation-tot}.

When the collisional processes are included, the trap decay is inhibited at earlier times because the resonant collisions provide a more efficient mechanism of increasing $\left| m\right|$ at short times. Overall, a better agreement with the experimental observations is reached at all $n$ values investigated, although the deviation between experimental and simulated curves increase with increasing $n$ values. The deviations between simulated and measured relative populations of trapped Rydberg atoms never exceed a factor of 1.5, which can be regarded as excellent given the simplicity of our collisional model. The simulations reveal that the collisions giving rise to the slow decay at late times take place at early times, when the atom density in the trap is high and the collisions are more likely. However, they only affect a small fraction of the atoms so that their effects are only observable once the population of trapped atoms has been reduced by a factor of more than ten.

Figure~\ref{fig:ch6-population-collision} displays the calculated distribution of the population of atoms in the trap over the different~$n$ (top panels), $k$ (middle panels) and $\left|m\right|$ (bottom panels) states at the time of photoexcitation ($t=0$) and at times $t=100\,\mu$s, 500\,$\mu$s, 1\,ms, 2\,ms and 3\,ms following Rydberg-Stark deceleration and trapping of initially excited $n=32$, $k=15-23$ Rydberg states.
These distributions were obtained by considering spontaneous emission only (black), all radiative processes (red), and all radiative processes and collisions (blue). The scales of the distributions calculated for the later times have been enlarged by the factors indicated in the upper right corners of the different panels. These factors are particularly large for the simulations that consider spontaneous decay only, because in this case, the only process by which atoms can remain trapped for long times involves the successive emission of several low-frequency photons in transitions that raise the value of $\left|m\right|$. This is reflected by a narrower range of populated $n$, $k$ and $\left|m\right|$ states than obtained with the simulations that also consider stimulated transitions and collisions. The simulations that consider all radiative processes but disregard the effects of collisions lead to the broadest distributions of $n$ and $k$ values. These distributions are the result of several successive stimulated absorption and emission processes. The $\left|m\right|$ distribution only broadens slowly so that the stabilization primarily results from transitions to states of significantly higher $n$ values. Because the successive transitions also broaden the distribution of $k$ values and increase the average value of $k$ (with a peak value at $k\approx 30$). The dipole moment of a Rydberg-Stark state is $\left|\vec{\mu}_{\text{el}}\right|\approx\frac{3}{2}nk\,a_0e$. Consequently, the radiative processes gradually increase the dipole moment of the trapped atoms, from an initial value of about 900 $a_0e$ to an average value of almost 1200 $a_0e$ after 3 ms. This increase corresponds to a raise of the translational temperature of the Rydberg sample in the electrostatic trap by a factor of $\sim 2$.

The collisions efficiently redistribute the population over a broad range of $\left|m\right|$ and $k$ values through the network of near-degenerate pair states of the type described by Eq.~(\ref{eq:15}) and which do not change $n$. In this case, the stabilization is the consequence of the increase in $\left|m\right|$ much more than of the increase of $n$. The distributions remain approximately constant after 1 ms, at which time they peak at $n_{\text{max}}=32$ and $k_{\text{max}}=19$, which is not significantly different from the initial distribution. The collisions thus prevent heating of the trapped sample. At early times ($t=100\,\mu$s), only states of even
$\left|m\right|$ values are populated in the trap, with $\approx 21\%$ $m=0$, $\approx 44\%$ $\left|m\right|=2$, $\approx 23\%$ $\left|m\right|=4$ and $\approx 6\%$ $\left|m\right|=6$ states, because of the intrinsic nature of the collisions considered in our model. Population of odd-$\left|m\right|$ states is initially induced by the blackbody-radiation field and becomes significant after about 500 $\mu$s. After 1 ms, the density of trapped hydrogen atoms in our experiments is so low that collision-induced transitions are negligible and the trap dynamics is entirely governed by radiative processes.
At 3\,ms, the distribution is centered around $\left|m\right|_{\text{max}}=12$ and a full width at half maximum $\Gamma_{\left|m\right|}\approx 12$.

The distribution of the population over several $n$ and a broad range of $m$ states reached after about 1 ms is similar to the distribution expected for the production of antihydrogen by positron transfer from positronium (see Ref.~\cite{kellerbauer08a} and references therein). The experimental results indicate a slow decay at this time, with a decay constant of (1 ms)$^{-1}$. We expect that this decay represents a realistic estimate of the decay dynamics after formation of antihydrogen following positron transfer from positronium. In view of trapping antihydrogen or of carrying out experiments with antihydrogen in its ground state, the rate at which the ground state is populated is of particular interest. Our results indicate that the relaxation time to the ground state should be about 1 ms for Rydberg states of principal quantum number around 30. This estimate is confirmed by Monte-Carlo simulations of the radiative processes assuming an initial population equally distributed over the states $n=30$, $\left|m_i\right|=0$-15. The result of the simulations presented in Fig.~\ref{aegis} indicate that, at 11\,K, about 65\% of the atoms are in the ground state after 1\,ms. To test how thermal radiation affects the fraction of ground state atoms, a simulation was also carried out for a 300\,K radiation field. Under these conditions, the decay of the initially prepared states is more rapid, but the yield of ground-state atoms at short times hardly increases because ionization competes with fluorescence to the ground state. It therefore appears that increasing the temperature of the radiation field does not significantly enhance the yield of ground-state atoms at any time. Instead, it reduces it at long times because more than half of the atoms are photoionized.

\begin{figure}[!tbh]
\centering
\includegraphics[angle=0,width=0.4\textwidth]{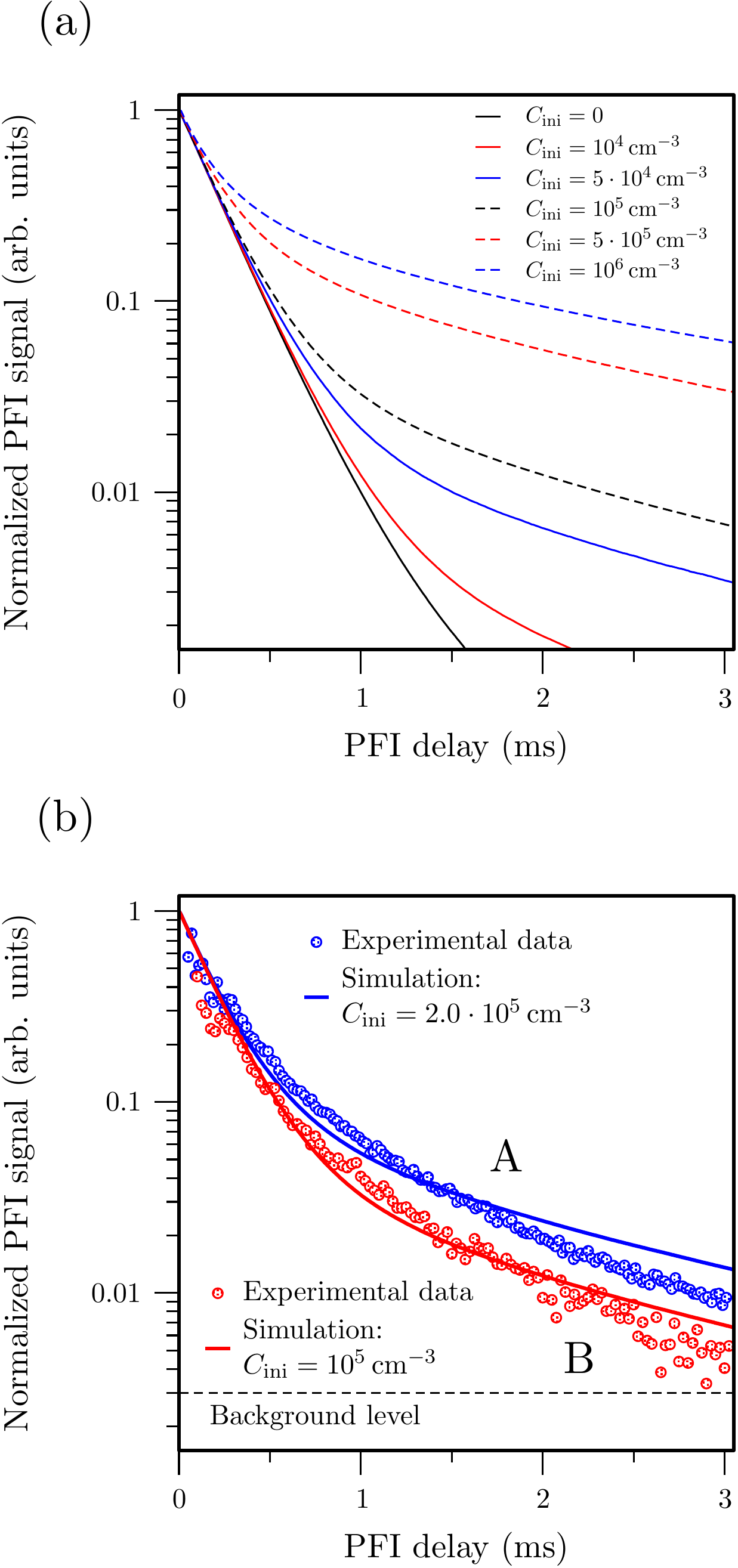}
\caption{\label{fig:ch6-n33-density}(a)~Simulations of decay of $n=33$ Rydberg hydrogen atoms from the trap for initial densities of $\leq 10^2$\,cm$^{-3}$ (black line), $10^4$\,cm$^{-3}$ (red line), $5\cdot 10^4$\,cm$^{-3}$ (blue line), $10^5$\,cm$^{-3}$ (dashed black line), $5\cdot 10^5$\,cm$^{-3}$ (dashed red line), and $10^6$\,cm$^{-3}$ (dashed blue line). (b)~Measurement of the H$^+$ ion signal obtained by pulsed-electric-field ionization as a function of the delay following photoexcitation to $n=33$ Rydberg-Stark states. The blue data points correspond to the experimental trap-decay curve already presented in Fig.~\ref{fig:ch6-simulation-comparison}. To obtain the red data points, the initial density was reduced by a factor of 2. The background level for this lower initial density is displayed as horizontal dashed black line. The experimental traces are compared with trap-decay curves simulated for particle densities of $N_0=2\cdot 10^5\,\text{cm}^{-3}$ (full blue trace labeled A) and $10^5\,\text{cm}^{-3}$ (full red trace labeled B).}
\end{figure}

Figure~\ref{fig:ch6-n33-density}(a) displays trap-decay curves calculated for $n=33$ Rydberg states assuming initial densities~$N_0$ of $10^4$\,cm$^{-3}$ (black line), $5\cdot 10^4$\,cm$^{-3}$ (red line), $10^5$\,cm$^{-3}$ (blue line), $5\cdot 10^5$\,cm$^{-3}$ (dashed black line) and $10^6$\,cm$^{-3}$ (dashed red line). At low initial densities ($N_0\leq 10^4\,\text{cm}^{-3}$), the separation between the trapped Rydberg atoms is much larger than $r_{\text{HS}}$, and collisions are negligible. Under these conditions, the trap decay curves are dominated by radiative processes and are therefore well described by the dashed lines in Fig.~\ref{fig:ch6-simulation-comparison}. Densities greater than $10^5\,\text{cm}^{-3}$ are large enough to induce $m$-changing collisions between trapped Rydberg H atoms with initial principal quantum numbers in the range $n=20-40$. These collisions increase the magnetic quantum number, an thus the radiative lifetimes of the trapped atoms. As a result, an enhanced pulsed field ionization signal is observed after $500\,\mu$s in experiments carried out at these initial Rydberg atom densities.

Comparison of the different curves depicted in Fig.~\ref{fig:ch6-n33-density}(a) enables us to estimate the initial density of atoms excited to Rydberg-Stark states to be about $N_0=2\cdot 10^5\,\text{cm}^{-3}$ in our experiments. This initial density corresponds within a factor of 2 to the initial density we estimate from the integrated H$^+$ signal of the time-of-flight spectra recorded following pulsed-electric-field ionization after a delay of $200\,\mu$s.
A reduction in the intensity of the second laser, which excites the intermediate $2\,^2\text{P}_{1/2,3/2}$ state to Rydberg-Stark states, enables one to reduce the initial density of Rydberg atoms in the trap. Fig.~\ref{fig:ch6-n33-density}(b) compares the measurement of the trap decay curve obtained following excitation to $n=33$ Rydberg-Stark states (blue data points) with a trap-decay curve recorded after reducing the initial density of Rydberg atoms by a factor of two (red data points). The corresponding simulated curves are displayed as full blue and red traces, respectively and reproduce the overall shape of the measured curves and the differences resulting from the reduction of initial density in a qualitative manner. We consider the agreement between the experimental and simulated trap decay curves depicted in Figs.~\ref{fig:ch6-simulation-comparison} and~\ref{fig:ch6-n33-density}(b) as evidence of the role played by collisions in our measurements of trap-decay curves and also as an indication that our simple model to treat the effects of these collisions captures the main aspects of the processes involved.


\section{\label{sc:conclusions}Conclusions}

The ability to trap samples of atoms in selected Rydberg-Stark states at low translational temperatures has been exploited to study the processes by which these atoms decay by radiative and collisional processes over extended periods of time, up to several milliseconds. Over such long times, sequences of radiative transitions, spontaneous and stimulated, and collisions between Rydberg atoms in the trap lead to distributions of states that are very different from the initially prepared ones. We have studied these processes experimentally by monitoring the rate at which the atoms are lost from the trap in dependence of the initial density of trapped atoms, the temperature of the radiation field and the principal quantum number, which was varied between 27 and 33. The decay of the atoms from the trap was found to be nonexponential and to depend on the principal quantum number and the temperature of the radiation field.

To interpret the experimental observations, Monte-Carlo simulations were performed based on the numerical evaluation of transition probabilities by spontaneous emission, absorption and emission stimulated by the thermal radiation, and also resonant collisions between Rydberg atoms.
The consideration, in these simulations, of radiative processes only, i.e., spontaneous emission at low temperatures and spontaneous and stimulated radiative processes included photoionization at room temperature, turned out to be sufficient to account for the near-exponential decay observed at early times, but underestimated the amount of atoms remaining in the trap by almost two order of magnitudes. The analysis of the decay curves observed experimentally and the comparison with the simulated decay curves indicated that only atoms having undergone transitions to Rydberg-Stark states of higher $\left|m\right|$ values remain in the trap after 1\,ms.

When looking for the origin of these transitions, we found that resonant collisions between two atoms in Stark states of the same $n$ manifold represent a more efficient mechanism to increase $\left|m\right|$ than radiative processes. Their efficiency comes from the strength of the dipole-dipole interaction and the fact that this interaction couples entire networks of near-resonant states of the interacting atoms at low fields, for which the Stark effect is linear. Indeed, the linearity of the Stark effect implies that if one atom undergoes a transition in which $k$ and $m$ change by $\pm 1$ and the other atom undergoes a transition in which $k$ and $m$ change by $\mp 1$ the energy remains constant.
To account for the effects of these transition networks induced by collisions, we have generalized the two-state F\"orster-type collision model used to describe resonant collisions between Rydberg atoms \cite{safinya81a,gallagher82a,pillet87a} and which plays a key role in the dynamics of frozen Rydberg gases \cite{anderson98a,mourachko98a,akulin99a} to a multi-state situation.

In analyzing the dynamics of the trapped Rydberg hydrogen atoms, we found that the efficiency of the collision-induced transitions is reduced by the inhomogeneous nature of the trapping electric field and derived a simple model to account for this reduction. Including collision-processes in the Monte-Carlo simulations enabled us to satisfactorily reproduce the experimental data by adjusting only the initial density of trapped atoms, the optimal value we found being $2\cdot 10^{5}$\,cm$^{-1}$, which is in agreement with the densities estimated from the magnitude of the pulsed-field-ionization signal and the volume of our trap.


\section*{Acknowledgments}

We thank H.~Schmutz for technical support, and Dr. S.~D.~Hogan (UCL) and S.~Hartweg (ETH Zurich) for fruitful discussions and their early contribution to this research. This work is supported financially by the Swiss National Science Foundation under Project 200020-132688, and by the NCCR QSIT.


\providecommand{\newblock}{}

\end{document}